\documentclass[twocolumn, times]{aastex631}
\usepackage{booktabs}
\usepackage{soul}
\usepackage{amsmath}
\usepackage{textcomp,gensymb}
\usepackage{placeins}

\newcommand\ha{H$\alpha$}
\newcommand\angstrom{\mbox{\normalfont\AA}}

\shorttitle{Deep SMFs of the Virgo cluster}
\shortauthors{Morgan et al.}

\begin{document}

\title{Deep galaxy stellar mass functions as a function of star formation rate in the Virgo cluster environment}

\author[0009-0009-2522-3685]{Cameron~R.~Morgan}
\affiliation{Waterloo Centre for Astrophysics, University of Waterloo, Waterloo, ON N2L 3G1, Canada}
\affiliation{Department of Physics and Astronomy, University of Waterloo, Waterloo, ON N2L 3G1, Canada}

\author[0000-0001-6245-5121]{Elizaveta~Sazonova}
\affiliation{Waterloo Centre for Astrophysics, University of Waterloo, Waterloo, ON N2L 3G1, Canada}
\affiliation{Department of Physics and Astronomy, University of Waterloo, Waterloo, ON N2L 3G1, Canada}

\author[0000-0002-0692-0911]{Ian~D.~Roberts}
\affiliation{Waterloo Centre for Astrophysics, University of Waterloo, Waterloo, ON N2L 3G1, Canada}
\affiliation{Department of Physics and Astronomy, University of Waterloo, Waterloo, ON N2L 3G1, Canada}

\author[0000-0003-4849-9536]{Michael~L.~Balogh}
\affiliation{Waterloo Centre for Astrophysics, University of Waterloo, Waterloo, ON N2L 3G1, Canada}
\affiliation{Department of Physics and Astronomy, University of Waterloo, Waterloo, ON N2L 3G1, Canada}

\author[0000-0002-0363-4266]{Joel~Roediger}
\affiliation{National Research Council of Canada, Herzberg Astronomy and Astrophysics Research Centre, Victoria, BC V9E 2E7, Canada}

\author[0000-0002-8224-1128]{Laura~Ferrarese}
\affiliation{National Research Council of Canada, Herzberg Astronomy and Astrophysics Research Centre, Victoria, BC V9E 2E7, Canada}

\author[0000-0003-1184-8114]{Patrick~C\^ot\'e}
\affiliation{National Research Council of Canada, Herzberg Astronomy and Astrophysics Research Centre, Victoria, BC V9E 2E7, Canada}

\author[0000-0000-0000-0000]{Alessandro~Boselli}
\affiliation{Aix-Marseille Univ., CNRS, CNES, LAM, Marseille, France}
\affiliation{INAF - Osservatorio Astronomico di Cagliari, via della Scienza 5, 09047 Selargius , Italy}

\author[0000-0002-9043-8764]{Matteo~Fossati}
\affiliation{Universit\'a di Milano-Bicocca, Piazza della Scienza 3, 20100 Milano, Italy}
\affiliation{INAF - Osservatorio Astronomico di Brera, via Brera 28, 21021 Milano, Italy}

\author[0000-0002-3263-8645]{Jean-Charles~Cuillandre}
\affiliation{AIM, CEA, CNRS, Universit\'e Paris-Saclay, Universit\'e de Paris, F-91191 Gif-sur-Yvette, France}

\author[0000-0001-8221-8406]{Stephen~Gwyn}
\affiliation{National Research Council of Canada, Herzberg Astronomy and Astrophysics Research Centre, Victoria, BC V9E 2E7, Canada}

\begin{abstract}

We analyze deep ($M_*\gtrsim10^7~{M}_{\odot}$) galaxy stellar mass functions (SMFs) of the Virgo cluster using stellar masses derived as part of the Next Generation Virgo Survey (NGVS). The total SMF has a slope of $\alpha=-1.35^{+0.02}_{-0.02}$ which is similar to or steeper than typical field values. Using deep \ha{} data from the Virgo Environmental Survey Tracing Ionised Gas Emission (VESTIGE) we separate out star-forming galaxies, quiescent galaxies with no ongoing star formation, and low-SFR galaxies that are intermediate between these two populations. For each of these populations, the shape of the SMF is found to be universal throughout the cluster, from the core to the outskirts. The star-forming and quiescent SMFs show stark differences with values seen in field galaxies. The relative fraction of quiescent galaxies is highest in the core of the cluster, with low-SFR and star-forming galaxies more significant in the outer regions of the cluster. At low stellar masses ($M_*\lesssim10^9~{M}_{\odot}$), the quiescent fraction in the main cluster is significantly higher than that of the field and even satellites of massive groups. At high stellar masses, the quiescent fraction is similar to other studies of cluster galaxies. We model the quiescent population in the infall region of the cluster as a combination of backsplash and field quiescent galaxies, and find that the backsplash fractions needed to explain the observed population are unrealistically high. This suggests the existence of a third population of low-mass galaxies that are pre-processed outside the virial radius of the cluster, possibly in groups prior to infall.

\end{abstract}


\section{Introduction} \label{sec:intro}

Galaxy stellar mass functions (SMFs) are a critical tool to aid in understanding the formation and evolution of galaxy populations across different environments and over cosmic time. In the local universe, the shape of the total galaxy SMF has been well-established through large spectroscopic surveys including the Two Micro Sky Survey (2MASS; \citealt{skrutskie2006}), the Sloan Digital Sky Survey (SDSS; \citealt{york2000, bell2003, baldry2006}), and the Galaxy And Mass Assembly survey (GAMA; \citealt{baldry2010, baldry2012}). When split into blue and red galaxy populations, the red population dominates in the high-mass regime, whereas blue galaxies dominate at lower stellar masses.

Observations of SMFs as a function of redshift suggest that massive galaxies form their stellar mass early on, with lower mass galaxies assembling later, in contrast with the theorized bottom-up growth of dark matter halos \citep[e.g.][]{faber2007, ilbert2013, muzzin2013, weaver2023}. With most of the stellar mass in the universe in place by $z \sim 2$, these studies show that the SMF of star-forming (SF) galaxies alone evolves only weakly to the present day \citep[e.g.][]{muzzin2013, mcleod2021}. The stagnancy of the SF SMF coupled with the buildup of massive, quiescent galaxies over time implies galaxies must move from SF to quiescent - galaxies ``quench'' through the cessation of star formation.

Studies that have measured local SMFs have sought to compare SMFs across a range of local densities or halo masses. In comparing SMFs as a function of local density, \citet{baldry2006} showed that the slope of the SMF shows little change with environment, whereas the characteristic mass increases with increasing density. \citet{yang2009} compared conditional SMFs (the SMF conditional on the halo mass $\mathcal{M}_h$) in SDSS groups with $10^{12}~\text{M}_{\odot} < \mathcal{M}_h < 10^{15}~\text{M}_{\odot}$. They found increasing characteristic mass and steeper low-mass slopes with increasing halo mass. More recently, a similar analysis was performed by \citet{vazquez2020} with the deeper GAMA survey. While the authors found the expected increase in characteristic mass with increasing halo mass, they found steeper low-mass slopes in the lowest halo mass bins, contrary to \citet{yang2009}.

While SMFs of galaxy clusters and proto-clusters have been studied at intermediate to high redshifts \citep[e.g.][]{vanderburg2020, edward2024}, SMFs in local ($D<100~\text{Mpc}$) massive clusters specifically have been less well studied, in part due to the large angles subtended on the sky. On average, cluster galaxies are more deficient in cold gas \citep{giovanelli1985, cayatte1990, gavazzi2005, cortese2011, boselli2014_hrs3, alberts2022} and are less actively star-forming \citep{balogh1997, balogh1998, lewis2002, gomez2003, kauffmann2004, baldry2006, weinmann2006, wetzel2012, gavazzi2013, boselli2023} than their field counterparts. \citet{balogh2001} analyzed the luminosity functions and corresponding SMFs of local field and cluster galaxies, finding that the cluster SMF has both a steeper low-mass slope and a larger characteristic mass. Analyzing the Shapley Supercluster, \citet{merluzzi2010} found a low-mass slope shallower than that of \citet{balogh2001}, and closer to observed values in the field. Recently, \citet{cuillandre2024} used early Euclid observations to measure the mass function of the Perseus cluster down to $10^7~\text{M}_{\odot}$, finding a shallower low-mass slope, similar to field values. These studies all show that the characteristic mass of the SMF is dependent on environment, with denser environments showing higher characteristic masses.

Recent observational work has indicated that galaxy quenching occurs primarily via one of two pathways: ``mass quenching'' and ``environmental quenching'', which have been shown to be separable in the local universe \citep{baldry2006, peng2010}. Mass quenching processes are stellar mass-dependent, internal mechanisms including supernova feedback \citep[e.g.][]{silk1998, hopkins2006} and active galactic nuclei (AGN) feedback \citep[e.g.][]{dekel1986, ceverino2009}. These two processes are effective at quenching low and high-mass galaxies, respectively. ``Environmental quenching'', then, refers to mechanisms that induce galaxy quenching externally. This includes hydrodynamical interactions such as ram-pressure stripping (RPS; \citealt{gunn1972, boselli2022}), where the interstellar medium of the galactic disk is stripped through interaction with the dense intracluster medium, rapidly removing star-forming material. Additionally, gravitational interactions can contribute to quenching in groups and clusters. Direct interactions such as mergers and tidal stripping can either strip material or induce starbursts by removing angular momentum from the gas in the disk, causing it to fall toward the centre \citep[e.g.][]{moore1996, moore1998}. Finally, galaxies can undergo starvation if they are unable to accrete or cool fresh gas. RPS can remove hot gas from the halo, or the cluster environment can prevent gas from cooling onto galactic disks \citep[e.g.][]{larson1980, balogh2000}.

The SMFs of the SF and quenched population as a function of environment and redshift can be used to trace the galaxy quenching process.   For large surveys, this is often done with colour-magnitude or colour-colour cuts. \citet{baldry2006} showed that, for SDSS galaxies, the increase in characteristic mass with increasing environmental density is due to the increase in massive, red galaxies in denser environments. Even where the shapes of the SMFs do not change considerably, the relative fraction of red galaxies increases with density across the stellar mass range. Interestingly, \citet{yang2009} found that when SDSS galaxies were split into blue and red samples, the SMFs of the two populations both showed a dependence on halo mass. Galaxies can be split into SF and quenched populations more directly using  direct tracers of star formation activity. \citet{balogh2001} split their sample into emission line and non-emission line galaxies, and noted that the shape of the SMF of cluster non-emitters is similar to the total field population (dominated by emission-line galaxies), implying the cluster population may be built up by field emission line galaxies that ceased star formation after cluster infall.

\begin{deluxetable*}{ccccccccc} 
\tablecaption{Properties of Virgo cluster substructures (taken from \citealt{boselli2023_lfha})}
\tablehead{\colhead{Substructure} & \colhead{RA} & \colhead{DEC}  & \colhead{Radius}  & \colhead{Vel range}  & \colhead{Distance}  & \colhead{$\langle v \rangle$}  &  \colhead{$\sigma$}   & \colhead{Central galaxy}\\[-0.20cm]
\colhead{} & \colhead{(deg)} & \colhead{(deg)}  & \colhead{(deg)} & \colhead{(km s$^{-1}$)} & \colhead{(Mpc)} & \colhead{(km s$^{-1}$)}   &  \colhead{(km s$^{-1}$)}  &  \colhead{}}\label{tab:structures}  
\startdata
Cluster A  &  187.71  &  12.39  & 5.383 & $<3000$ &  16.5 &  955 &  799 & M87  \\
Cluster B  &  187.44  &  8.00  & 3.334 & $<3000$ &  16.5 &  1134 &  464 & M49  \\
Cluster C  &  190.85  &  11.45  & 0.7 & $<3000$ &  16.5 &  1073 &  545 & M60  \\
W Cloud  &  185.00  &  5.80  & 1.2 & $ 1000 < v <3000$ &  32 &  2176 &  416 & NGC~4261   \\
W$'$ Cloud &  186.00  &  7.20  & 0.8 & $<2000$ &  23 &  1019 &  416 & NGC~4365  \\
M Cloud  &  183.00  &  13.40  & 1.5 & $1500 < v <3000$ &  32 &  2109 &  280 & NGC~4168  \\
LV Cloud  &  184.00  &  13.40  & 1.5 & $<400$ &  16.5 &  85 &  208 & NGC~4216   
\enddata
\end{deluxetable*}

What is missing from the literature is a comprehensive analysis of SMFs in the local Universe down to low stellar mass across local environments, with a robust star formation indicator. It is clear that the distribution of stellar mass depends on environment whether looking at the total SMF or the SMFs of SF and quiescent galaxies separately. However, it remains unclear how the populations in cluster infall regions differ from the cluster core, for example. Further, pre-processing of galaxies in smaller halos before infall into a massive cluster is likely an important yet poorly constrained step in the environmental evolution of galaxies \citep[e.g.][]{fujita2004}.

The Virgo cluster is an ideal environment to study the effects of environmental quenching at low-redshift. Virgo is a dynamically unrelaxed cluster. While it is dominated by a main cluster (Cluster~A) centred around M87, a second smaller cluster (Cluster~B, centred around M49; \citealt{mei2007}) is merging into Cluster~A from the south. Additionally, several smaller substructures have been identified, some of which appear to be infalling toward Virgo along filaments \citep{gavazzi1999, mei2007, cantiello2024}. Thus, the Virgo environment contains galaxies in various stages of infall into the cluster, including populations in small structures that have not yet felt the influence of the main cluster.

At a distance of 16.5~Mpc \citep{mei2007}, it is possible to image Virgo cluster galaxies at high physical resolution. The Next Generation Virgo Survey (NGVS; \citealt{ferrarese2012}) has done just this in the optical with a deep survey of the Virgo environment complete to $M_* \sim 10^7~\text{M}_{\odot}$. More recently, the Virgo Environmental Survey Tracing Ionised Gas Emission (VESTIGE; \citealt{boselli2018}) followed up NGVS with a blind, narrowband \ha{} survey. The combination of these two surveys provide optical and star-formation maps at resolutions $<100~\text{pc}$ for galaxies in the Virgo environment spanning six decades in stellar mass. Numerous studies of the effects of the Virgo environment on galaxy evolution have been produced using this data \citep[e.g.][]{ferrarese2016, roediger2017, lim2020, fossati2018, boselli2021, boselli2023, morgan2024}.

In this work, we take advantage of the depth of NGVS and VESTIGE to probe the SMFs within the Virgo environment. With the comprehensive \ha{} data from VESTIGE, we can isolate galaxies into star-forming, transition and quiescent populations without relying on colour as a proxy for star-formation as often done in the literature. While SMFs have been used to understand galaxy populations in the field at low-redshift \citep[e.g.][]{baldry2006, baldry2012}, and clusters and protoclusters at intermediate and high redshifts \cite[e.g.][]{vanderburg2020, edward2024}, few studies have comprehensively analyzed low redshift cluster environments using this technique. Here we compare the SMFs of quiescent, transition and SF galaxies across the Virgo environment at a depth not before studied. With this effort, we look to understand the origin of the environmentally quenched population in Virgo as a function of stellar mass. This work represents the most comprehensive study of the stellar mass distribution throughout a local cluster, down to low stellar mass and using a robust star formation indicator.

\section{Data} \label{sec:data}

\subsection{NGVS} \label{sec:NGVS}

The Next Generation Virgo Survey (NGVS; \citealt{ferrarese2012}) includes optical imaging of the Virgo cluster environment out to 1.55~Mpc from M87, and 0.96~Mpc from M49. Using MegaCam on the Canada-France-Hawaii Telescope (CFHT), the survey covered 104~deg$^2$ in the $u^*$, $g$, $i$ and $z$ bands (with partial coverage in the $r$ band). The long exposures of the survey allow for $5\sigma$ point-source depths of 26.3, 26.6, 25.8, and 24.8~mag in each of the four complete bands, respectively, with surface brightness depths of 29.3, 29.0, 27.4 and $26.0~\text{mag arcsec}^{-2}$.

MegaCam on  the CFHT consists of an array of 40 CCDs with pixel scale of 0.187$^{\prime\prime}~\text{pixel}^{-1}$. Seeing conditions during observations allowed for a resolution of $<1^{\prime\prime}$ (full-width-half-maximum). NGVS images were processed using Elixir-LSB, a pipeline specifically developed for enhanced detection of low-surface brightness features in NGVS images. Photometric calibration and astrometric calibration follows procedures detailed in \citet{gwyn2008}.

\begin{figure}[ht!]
\centering
\includegraphics[width=\columnwidth]{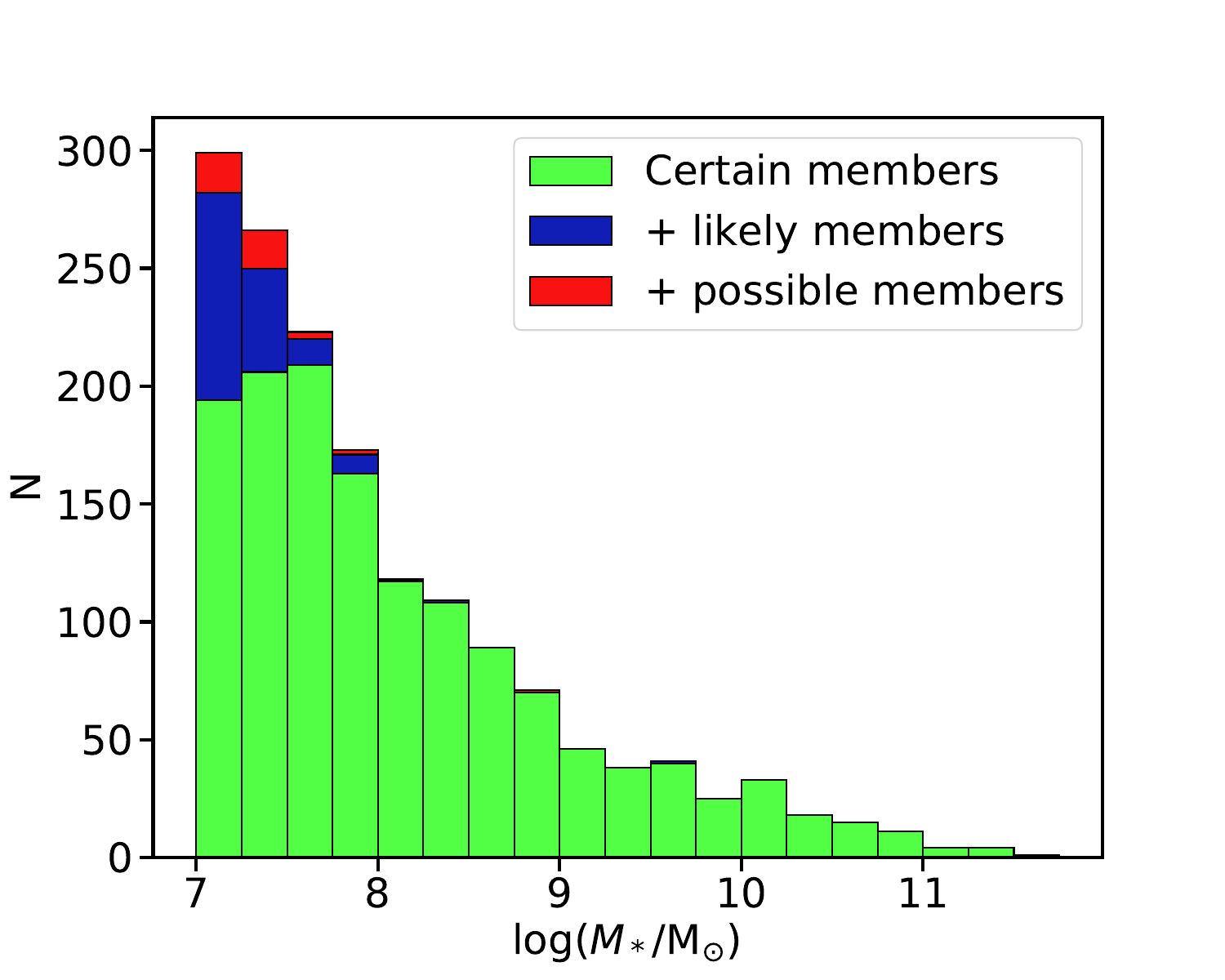}
\caption{Virgo cluster members binned by stellar mass based on the three different groups of possible cluster members.}
\label{fig:members}
\end{figure}

\subsection{VESTIGE} \label{sec:VESTIGE}

The Virgo Environmental Survey Tracing Ionised Gas Emission (VESTIGE; \citealt{boselli2018}) is a blind, narrowband \ha{} followup to NGVS, covering the same 104~deg$^2$ of the Virgo cluster using the narrowband MP9603 filter on CFHT MegaCam ($\lambda_c=6591~\angstrom $; $\Delta\lambda = 106~\angstrom$) as well as the broadband $r$ filter. The \ha{} Balmer line at $6563~\angstrom$ and the two [N\textsc{ii}] lines at $6548~\angstrom$ and $6583~\angstrom$ fall within the narrowband filter at the redshift range of Virgo ($-300 < v_{\rm hel} < 3000~\text{km s}^{-1}$). Long integrations (2h) in the narrowband achieved depths of $f(\text{H}\alpha)\simeq 4 \times 10^{-17}~\text{erg s}^{-1}~ \text{cm}^{-2}$ at $5\sigma$ for point sources and $\Sigma(\text{H}\alpha)\simeq 2 \times 10^{-18}~\text{erg s}^{-1}~ \text{cm}^{-2}~ \text{arcsec}^{-2}$ ($1\sigma$ after smoothing the data to $\sim \! 3''$ resolution) for extended sources. $r$ band exposures were 12~minutes, achieving a point source depth of $24.5~\text{mag}$ ($5\sigma)$ and a SB limit of $25.8~\text{mag arcsec}^{-2}$ ($1\sigma$ for scales comparable to the size of structures in target galaxies, $\sim30~\text{arcsec}$). Observations were taken under excellent seeing conditions (typical full-width-half-maximum of $0.76^{\prime\prime} \pm 0.07^{\prime\prime}$). At the distance of Virgo, $1^{\prime\prime}$ is equal to $80~\text{pc}$. The full observing strategy is discussed in detail in \citet{boselli2018}.

\subsection{Cluster membership} \label{sec:member}

To construct a robust catalogue of Virgo cluster members down to very faint luminosities and for which spectroscopic information was not available, the NGVS team developed a pipeline that included diagnostic plots based on photometric measurements, cuts on galaxy size, membership probability, and visual inspection. The result was a catalogue of galaxies with membership probability given a 0 (certain members), 1 (likely members) or 2 (possible members). The full strategy is discussed in detail in \citet{ferrarese2020} where this process was applied to the initial catalogue for the core of the cluster. The catalogue used in this work extends to the rest of the NGVS sample. We show in Figure~\ref{fig:members} a histogram of the number of galaxies assigned each of these membership probabilities as a function of stellar mass. Across most of the stellar mass range, the difference when ``likely'' and ``possible'' members are included is negligible, and there is only a small increase in the number of galaxies at very low stellar masses if ``possible members'' are included. We choose to only include ``certain'' and ``likely'' members for this study. Including ``likely'' members ensures a complete sample, but one that may be impure at the lowest stellar masses. However, we perform the forthcoming analysis with and without including ``likely'' members and find no significant difference in any of the results.

\section{Methods} \label{sec:methods}

\subsection{Stellar masses} \label{sec:stellar_masses}

Stellar masses were derived as part of the NGVS project, as detailed in Roediger et al. (in prep). To summarize, spectral energy distribution (SED) fitting was performed on the $u^*griz$ photometric data measured within one effective radius. All galaxies were assumed to be at the distance of the centre of Virgo, 16.5~Mpc. Flexible Stellar Population Synthesis simple stellar populations \citep{conroy2009} were used with a Chabrier initial mass function \citep{chabrier2003} and an exponentially-declining star-formation history. Mass-to-light ratios were then derived by fitting the observed SED to a grid of 50,000 synthetic models with a range of metallicities and ages. The final stellar mass catalogue is complete to $1.5\times10^7~\text{M}_{\odot}$, or $\mathcal{M}_{lim}=\log{\left( M_{*, \rm lim}/\text{M}_{\odot} \right)}\sim 7.2$, which we adopt as our mass limit.

\begin{figure}[ht!]
\centering
\includegraphics[width=\columnwidth]{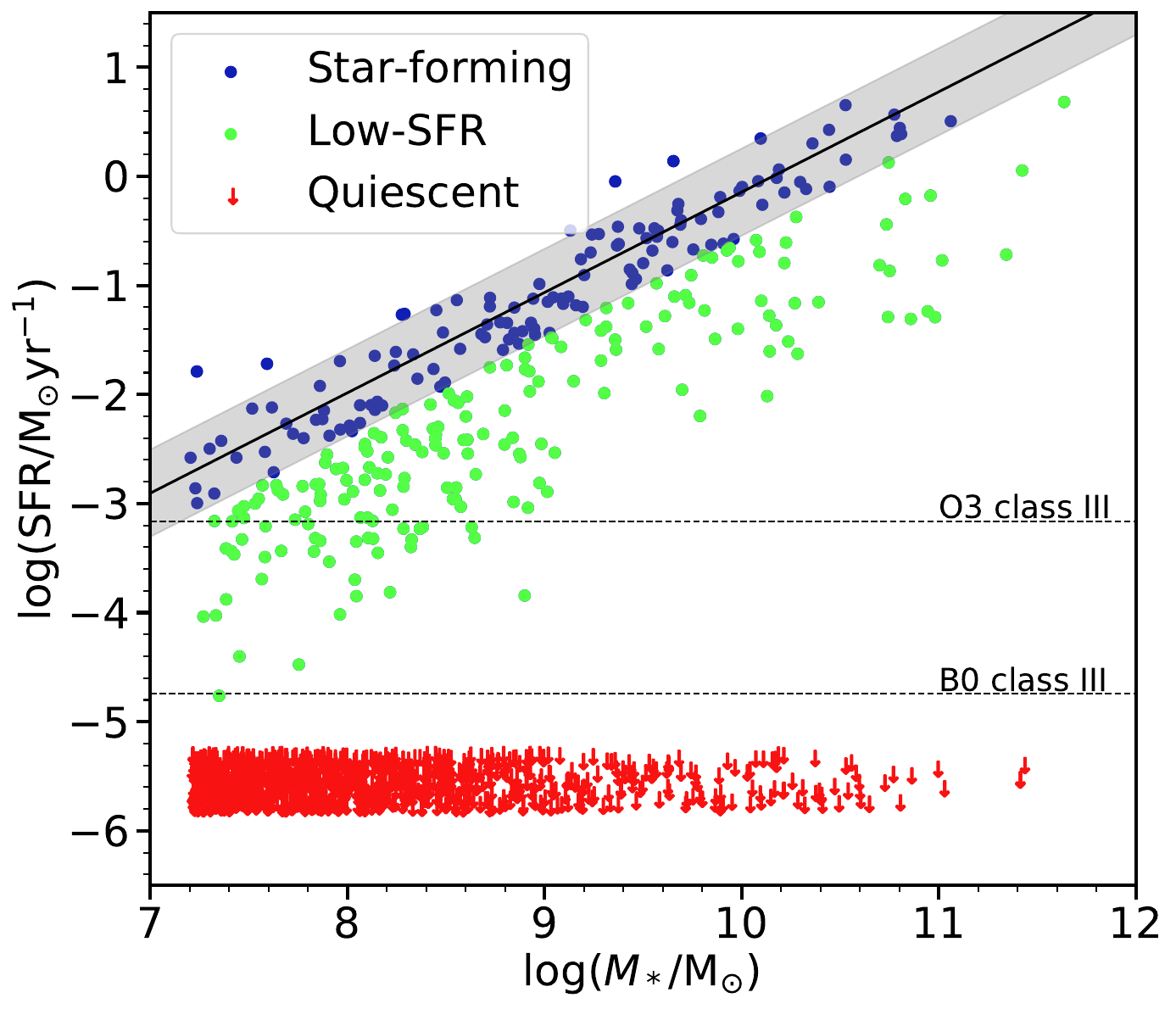}
\caption{SFMS of the VESTIGE sample with the main sequence defined as in \citealt{boselli2023}. Galaxies within $1\sigma$ of the SMFS are considered SF galaxies (blue), while those detected in \ha{} but with SFRs that place them below the SFMS are shown as low-SFR galaxies (green). Galaxies undetected in \ha{} are quiescent galaxies (red), and placed at random low values of SFR for illustration only.}
\label{fig:SFMS}
\end{figure}

\subsection{Subcluster membership and mass scaling}

\label{sec:sub_scale}

The Virgo cluster is made of several different substructures, in addition to the main cluster, Cluster A, centred around M87 \citep{boehringer1994}. The most prominent of these additional structures is Cluster B, centred on M49 \citep{gavazzi1999, mei2007}. Some works, based on the analysis of \citet{gavazzi1999} have used a distance of 23~Mpc for Cluster B. However, we follow studies that have placed Cluster B at similar distance to Cluster A \citep[e.g.][]{mei2007, cantiello2024}. In addition, Cluster C and the Low-Velocity (LV) Cloud are small structures identified within the footprint of and at the same distance as Cluster A. The W, W$^\prime$ and M Clouds are smaller structures that have been determined to lie behind the Virgo cluster, falling toward the main cluster body along filaments \citep{mei2007, cantiello2024}. Following \citet{mclaughlin1999}, \citet{ferrarese2012} determined $R_{200}$ for Clusters A and B, calculating 1.55~Mpc and 0.96~Mpc, respectively. These values were used to establish the footprint for NGVS and VESTIGE, and are listed as the associated radii for these structures in Table~\ref{tab:structures}. However, more recent X-ray studies have suggested that the virialized region of the main cluster is smaller, $R_{200}=0.974~\text{Mpc}$ \citep{simionescu2017}. In our analysis, we will use this value for $R_{200}$ of Cluster A and discuss the impact of this choice in context.

\begin{figure}[ht!]
\centering
\includegraphics[width=\columnwidth]{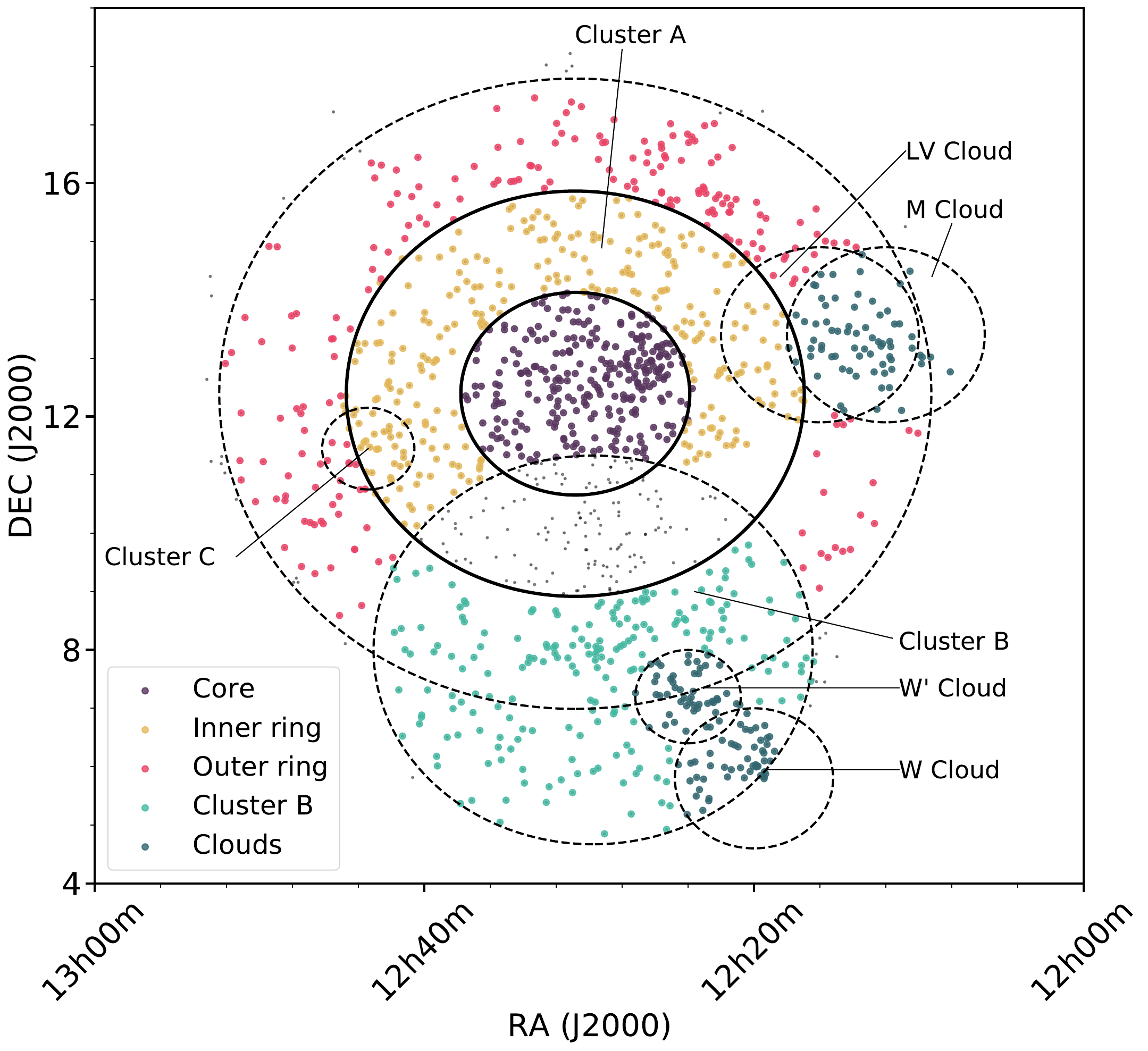}
\caption{RA and DEC footprint of galaxies in the Virgo sample. Dashed circles outline the various substructures associated with the Virgo cluster. The solid black lines at 0.5 and 1 Mpc define boundaries for the three radial bins in which galaxies associated with Cluster A are grouped. Galaxies are coloured based on the bin they belong to, with smaller grey data points not belonging to any bin.}
\label{fig:footprint}
\end{figure}

Galaxies can be assigned to their most likely substructure based on 2D position and radial velocity information. However, our sample is only complete with radial velocities above $\sim 10^{8.5}~\text{M}_{\odot}$. As such, we assign galaxies to substructures based on 2D position only with the goal of keeping the Cluster A sample as pure as possible. Galaxies are preferentially assigned to the W, W$^\prime$ and M Clouds if they fall within the footprint of these structures. Galaxies are assigned to Cluster B if they are located within the footprint of that structure and are $>1~\text{Mpc}$ ($\sim1R_{200}$) from M87 at the centre of Cluster A. The remaining galaxies all lie within the footprint of Cluster A and may also be part of Cluster C and the LV Clouds. Since these structures are small, contain few galaxies and are located within Cluster A, we include these galaxies in Cluster A. The five populations discussed are shown in Figure~\ref{fig:footprint} amongst the footprint of the NGVS survey. The radii used to define the boundaries of each substructures are listed in Table~\ref{tab:structures}.

Our adopted distances for each structure are also shown in Tab.~\ref{tab:structures}. Because the distance used in the calculation of the stellar masses was 16.5~Mpc for all galaxies, the stellar masses for galaxies in the W, W$^\prime$ and M Clouds will be underestimated. To account for this, we scale the masses based on the distances to these structures. Since we initially made a stellar mass cut at $\mathcal{M_{\rm low}}=7.2$, scaling the masses according to distance will move this lower limit to $\mathcal{M^\prime_{\rm lim}}=7.78$, since $\mathcal{M^\prime_{\rm lim}}=\mathcal{M_{\rm lim}} +2\log{\left(d_{\rm max}/16.5\right)}$, where $d_{\rm max}=32~\text{Mpc}$ is the distance to the furthest cloud and 16.5~Mpc is the distance to the main Virgo cluster.

\subsection{Star formation rates}

Integrated star-formation rates for all VESTIGE galaxies were derived in \citet{boselli2023}, and we refer the reader to this work for a full description of the process and considerations made. In brief, SFRs were derived using the \citet{calzetti2010} calibration converted to a Chabrier initial mass function (IMF). This produces the following relation: 

\begin{equation} 
    \text{SFR}\left[M_{\odot}~\text{yr}^{-1} \right] = 5.01 \times 10^{-42} L\left(\text{H}\alpha \right)\left[\text{erg s}^{-1}   \right]. 
\end{equation}

As discussed in \citet{boselli2023}, the sensitivity of VESTIGE probes \ha{} luminosities low enough to detect ionizing radiation produced by a single O or B-type star. At this extreme low range, effects such as stochasticity in the IMF can create large uncertainties on the SFR. In this work, we only consider galaxies above $10^7~\text{M}_{\odot}$, where we expect at least galaxies on the star-forming main sequence will have well-measured SFRs.

\citet{boselli2023} assigned galaxies to the various substructures of Virgo based on 2D position and radial velocity measurements. When calculating the SFRs, each galaxy was assumed to be at the distance corresponding to its assigned substructure. Since our substructure assignments do not completely overlap with those of \citet{boselli2023}, we scale SFRs where necessary based on the distance assigned to a galaxy (and its substructure) as described in Section~\ref{sec:sub_scale}. This ensures the stellar mass and SFR for each galaxy are based on the same distance. 

Using the measured SFRs, the VESTIGE SFMS was derived and analyzed in detail in \citet{boselli2023}. The best fit to the sample of H\textsc{i}-normal galaxies is 

\begin{equation}
\log{(\text{SFR})} = 0.92\log{(M_*)} - 9.34,
\end{equation}

\noindent with a scatter of $\sim 0.4~\text{dex}$. The SFMS is defined as a band spanning $\pm 1\sigma$ of the best fit line. We show this fit with our sample of galaxies plotted on Figure \ref{fig:SFMS}. 118 galaxies fall on the SFMS, with 8 galaxies located above the SFMS. For the purposes of this analysis, these galaxies with excess star formation are grouped into the SF sample. Thus, there are 126 galaxies in the final SF sample. Any galaxies with no detectable \ha{} emission are considered quiescent galaxies.

Massive early-type galaxies with AGN may have \ha{} emission that is not associated with star formation. \citet{boselli2023} checked for AGN contamination from nuclear spectra that is nearly complete for the VESTIGE sample \citep{cattorini2023}, finding that the handful of galaxies that may have AGN based on BPT \citep{baldwin1981} or WHAN \citep{fernandes2011} classifications all had clearly identifiable H\textsc{ii} regions dominating their \ha{} emission when visually inspected. As such, we have not removed any galaxies or altered their measured SFRs due to AGN contamination.

A large sample of galaxies have measurable SFRs that are $>1\sigma$ below the SFMS. We consider these galaxies to be low-SFR galaxies, and isolate them from our normal SF population for further analysis. A subset of low-SFR galaxies may represent a transition population between SF and quiescent galaxies. However, emission driven by low-luminosity AGN or LINERs cannot be completely ruled out. In the literature, the population is often defined using colour-colour cuts and may consist of transition galaxies and/or dust-obscured disks \citep[e.g.][]{schawinski2014}. However, we are able to directly measure the SFRs of galaxies and thus probe this transition population more directly. Recently, \citet{boselli2023} noted that these transition galaxies are mainly H\textsc{i}-deficient systems, and that their properties were consistent with RPS. However, work by \citet{morgan2024} suggested that the significant fraction of Virgo cluster galaxies in this transitionary phase could be indicative of slower quenching modes such as starvation, particularly for galaxies located beyond the cluster virial radius.

\begin{figure*}[ht!]
\centering
\includegraphics[width=2\columnwidth]{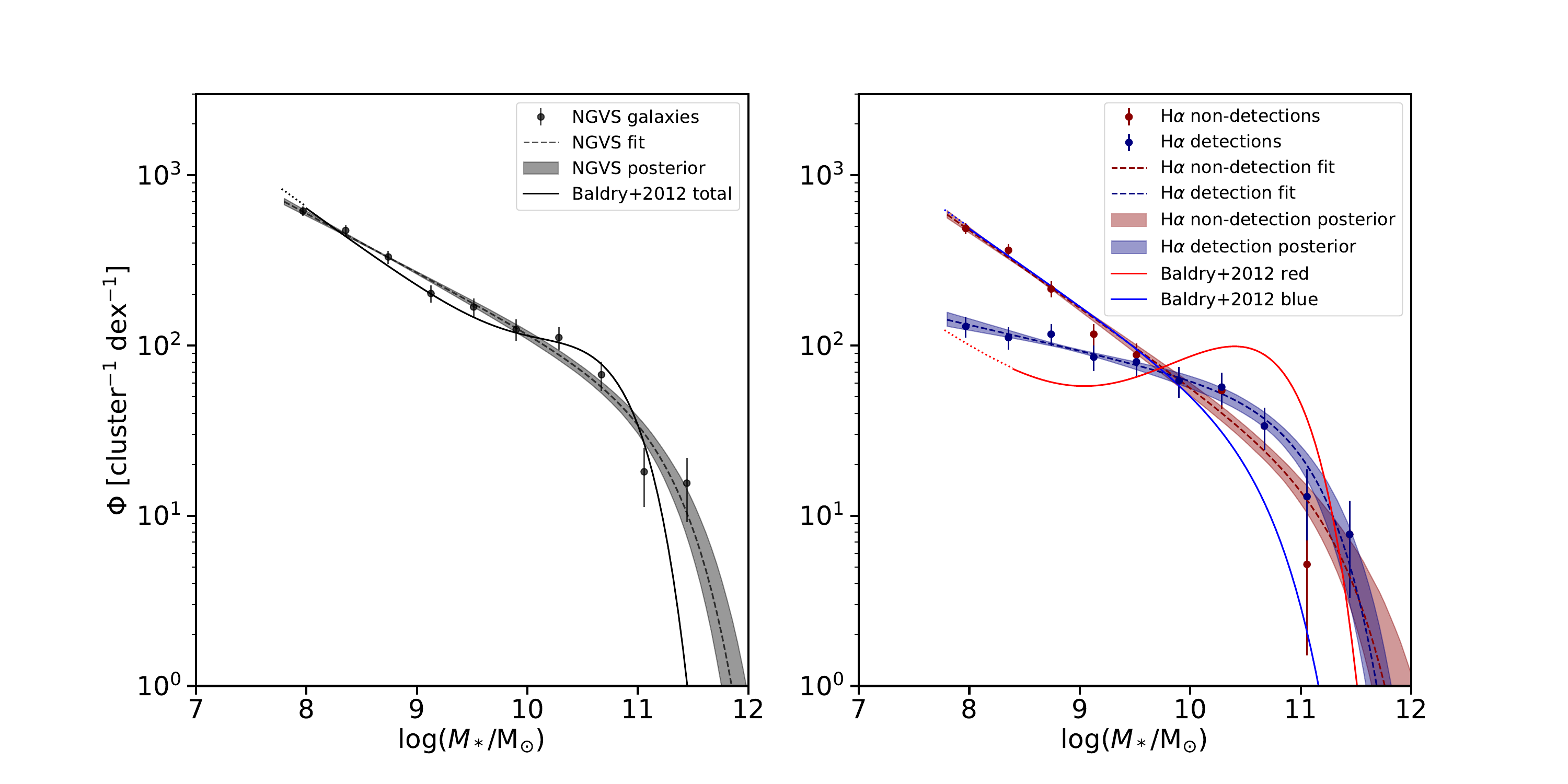}
\caption{SMFs of the entire Virgo sample compared to the field from GAMA. \textit{Left:} The total SMF of the Virgo sample is shown with the renormalized field SMF from GAMA \citep{baldry2012}. \textit{Right:} The SMFs of the \ha{} (blue) and non-\ha{} (red) selected populations in Virgo. The renormalized blue and red SMFs from the GAMA field are shown for comparison. The data points and errorbars (determined from Poisson statistics)} show binned data for visualization. The single Schechter is fit to the unbinned data, with the best fit shown along with the $1\sigma$ uncertainties. The dotted extensions of the GAMA curves are extrapolations below the mass limits of the survey.
\label{fig:smf_full}
\end{figure*}

\subsection{Stellar mass functions} \label{sec:smf}

We fit the distribution of logged stellar masses ($\mathcal{M}=\log{\left(M_*/\text{M}_{\odot}\right)}$) using a Schechter function \citep{schechter1976}, taking the form:

\begin{equation} \label{eq:schecter}
\begin{split}
    \phi(\mathcal{M}) = \text{ln}(10) \phi^{*} \cdot  \text{exp}(-10^{(\mathcal{M} - \mathcal{M}^{*})})
    \cdot10^{(\mathcal{M} - \mathcal{M}^{*})(\alpha+1)},
\end{split}
\end{equation}

\noindent where $\mathcal{M}^* = \log_{10}{(M^*/\text{M}_{\odot})}$ is the characteristic mass, $\alpha$ is the low-mass slope of the distribution, and $\phi^*$ is the normalization. We fit the Schechter function to the un-binned mass data using the parametric maximum-likelihood method \citep{sandage1979}. To do so, we assign a probability to each mass in the sample in the manner presented in \citet{oegerle1986} and \citet{malumuth1986}, where

\begin{equation} \label{eq:prob}
    p(\mathcal{M}_i) \equiv \frac{\phi(\mathcal{M}_i)}{\int_{\mathcal{M}_{\rm lim}}^{\infty} \phi(\mathcal{M}') \, d\mathcal{M}'}.
\end{equation}

\noindent The log-likelihood is simply the sum of the logarithms of the individual probabilities. By maximizing the log-likelihood we estimate the posterior distribution of the parameters $\mathcal{M}^*$ and $\alpha$ using the Python package \textsc{emcee} \citep{emcee}. The normalization $\phi^*$ is not fit as a free parameter, but is calculated such that the integral of the fit Schechter function over the entire mass range of the sample is equal to the number of galaxies in the sample.

Several studies analyzing SMFs have shown that a double-Schechter fit can better represent the mass distribution of galaxy samples than a single-Schechter \citep[e.g][]{baldry2008}. The double-Schechter function used is simply a two-term function where each term has the same form as Eq. \ref{eq:schecter}. The $\alpha$ and $\phi^*$ parameters for each term are fit, while the $\mathcal{M}^*$ parameter is fit but is the same in both terms. This functional form allows for cases such as a positive power law near the exponential cutoff, with an upturn (negative power law) to lower stellar masses. We performed double-Schechter fits on our total and quiescent populations where the binned data appeared visually to follow this shape. However, a Bayesian Information Criterion (BIC) test showed that the single-Schechter is statistically a better fit to the data, given the risk for overfitting with additional parameters. As such, we use exclusively single-Schechter fits in our analysis.

\begin{deluxetable*}{cccccc}
\tablecaption{Best fit parameters and uncertainties for the total Virgo SMF and the SMFs split according to \ha{} emission}
\tablehead{\colhead{Environment}&\colhead{Sample}&\colhead{N}&\colhead{$\alpha$}&\colhead{$\mathcal{M}^*$}&\colhead{$\phi^*$}\\[-0.20cm]
\colhead{}&\colhead{}&\colhead{}&\colhead{}&\colhead{$(\log{(M_*/\rm{M}_{\odot})})$}&\colhead{$(\rm{cluster}^{-1}\rm{dex}^{-1})$}}\label{tab:sfms_params}
\startdata
 & Total & 1308 & $-1.35^{+0.02}_{-0.02}$ & $11.33^{+0.17}_{-0.11}$ & $18.25^{+3.54}_{-4.53}$ \\
 & non-H$\alpha$/Quiescent & 997 & $-1.44^{+0.02}_{-0.02}$ & $11.35^{+0.31}_{-0.13}$ & $6.88^{+1.68}_{-2.46}$ \\
All & H$\alpha$ & 269 & $-1.15^{+0.03}_{-0.04}$ & $11.13^{+0.16}_{-0.11}$ & $19.47^{+4.07}_{-4.51}$ \\
 & Low-SFR & 158 & $-1.23^{+0.04}_{-0.05}$ & $11.45^{+0.35}_{-0.16}$ & $6.13^{+1.70}_{-2.47}$ \\
 & Star-forming & 111 & $-0.96^{+0.06}_{-0.07}$ & $10.60^{+0.17}_{-0.12}$ & $21.12^{+5.10}_{-5.65}$
\enddata
\end{deluxetable*}

\section{Results} \label{sec:results}

We present SMFs for the total Virgo environment, as well as SMFs split into environmental bins. Further, we break up our sample based on SFR determined from VESTIGE \ha{} data. We have removed the brightest cluster galaxies of Clusters A and B (M87 and M49, respectively) from our sample, such that our SMFs are of satellite galaxies of the cluster. We have not removed the central galaxies from the smaller substructures listed in Table~\ref{tab:structures}. We show the fit parameters for the SMFs of the full Virgo sample in Table~\ref{tab:sfms_params}. In Appendix~\ref{appendix}, Table~\ref{tab:sfms_params_full} displays the results from fits to individual environmental bins.

\begin{figure*}[ht!]
\centering
\includegraphics[width=2\columnwidth]{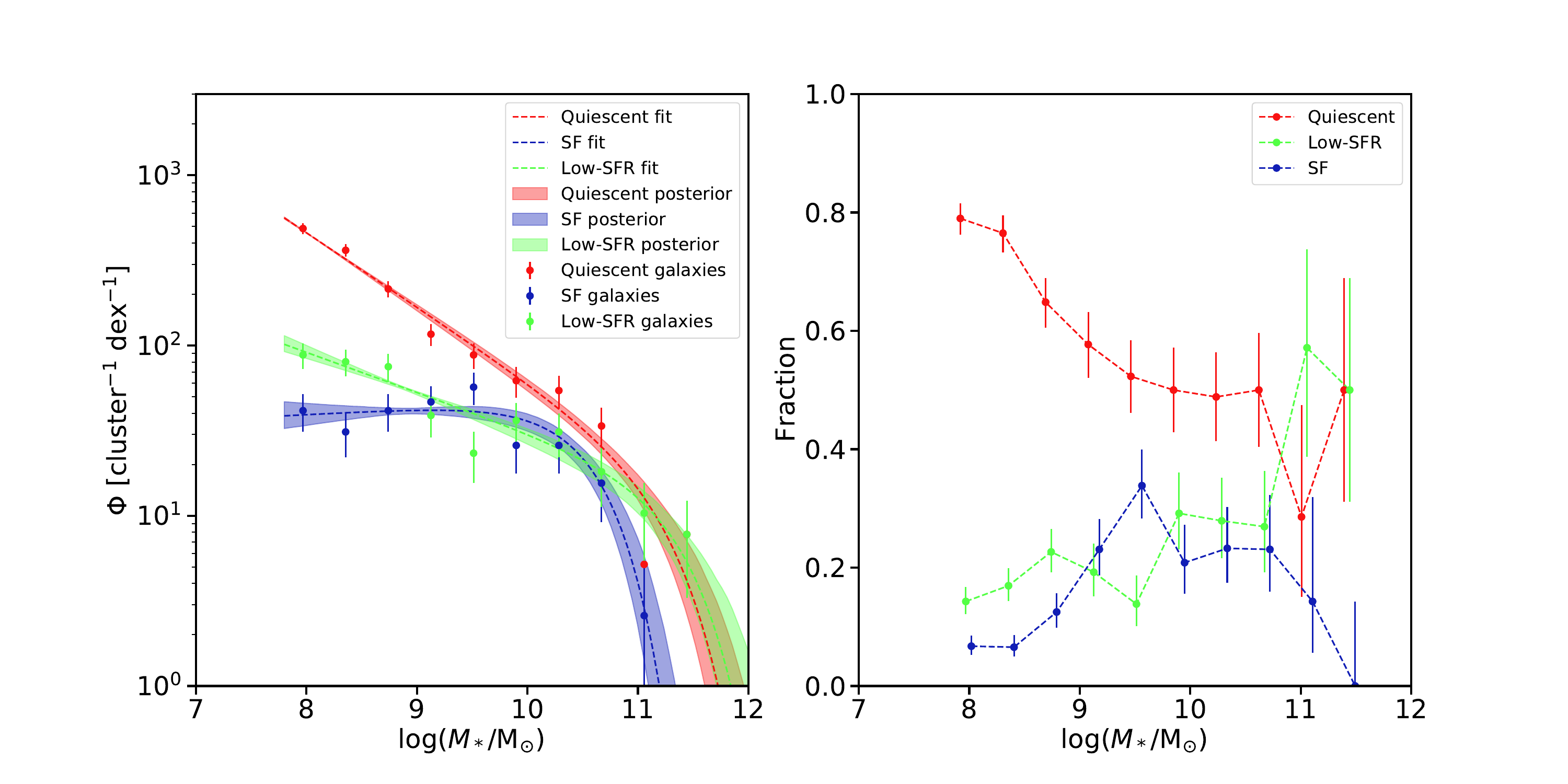}
\caption{SMFs and relative fractions of different populations (quiescent, low-SFR and SF) in Virgo. \textit{Left:} The SMF of quiescent (red), SF (blue) and low-SFR (green) populations in the Virgo sample. The data points and errorbars (determined from Poisson statistics)} show binned data for visualization. The single Schechter is fit to the unbinned data, with the best fit shown along with the $1\sigma$ uncertainties. \textit{Right:} The fraction of quenched (red), low-SFR (green), and SF (blue) galaxies in the Virgo sample as a function of stellar mass, with associated uncertainty ranges.
\label{fig:smf_sfms1}
\end{figure*}

\subsection{Full Virgo stellar mass functions} \label{sec:virgo_smf}

We fit the full Virgo SMF with a single Schechter fit as described in Section \ref{sec:methods} for $\mathcal{M'}_{lim} > 7.78$. Because the total mass function includes galaxies in the W, W$^\prime$, and M Clouds where we have scaled the masses, we only fit the total mass function to $\mathcal{M'}_{lim}$, since our sample of stellar masses will be incomplete below this mass. We show the results on the left panel of Figure \ref{fig:smf_full}, including the $1\sigma$ posterior distribution. For a comparison with a local field population, we show the fit from \citep{baldry2012} to field galaxies in the GAMA survey. While the GAMA survey completeness extends down to only $\mathcal{M} > 8$, we have renormalized the fit such that the integral of the field SMF is equal to the total number of galaxies within the mass range of the Virgo sample.

Since we will use the \ha{} data from the VESTIGE survey to separate Virgo galaxies into star-forming and quiescent populations, we separate galaxies into \ha{} detections and non-detections on the right side of Figure \ref{fig:smf_full}. As we expect both VESTIGE and NGVS samples to be complete down to $\mathcal{M'}_{lim} > 7.78$, the \ha{}-selected SMF contains all cluster galaxies detected with VESTIGE above this limit. The non-\ha{} SMF, then, contains all NGVS Virgo detections without an \ha{} counterpart in VESTIGE.  For comparison to the field, we plot the SMFs from \citet{baldry2012} separated into blue and red populations. The red and blue populations were separated using a $u-r$ colour-magnitude cut, and the SMFs for the two populations fit simultaneously. We have renormalized the \citet{baldry2012} curves such that the sum of the integrals under the blue and red curves is equal to the total number of galaxies in the Virgo sample, but the relative fractions of blue and red galaxies follow the GAMA field values.  We note that \citet{gunawardhana2015} presented the mass function of \ha{}-selected galaxies in the GAMA dataset. While this selection is a better match to ours, their \ha{}-selected SMF is in fact very similar to the \citet{baldry2012} blue SMF. As such, we plot the blue and red populations for comparison here and discuss them later in the paper.

\subsection{Separating populations based on star formation rate} \label{sec:sf_q}

On the left side of Figure \ref{fig:smf_sfms1} we show the SMFs of the quiescent, SF and low-SFR populations in Virgo. Each SMF was fit with a single Schechter function as described in Sect. \ref{sec:methods} for $\mathcal{M'}_{lim} > 7.78$. On the right side we show the quiescent, SF and low-SFR fractions as a function of stellar mass. We use the binned stellar masses to calculate the relative fractions and uncertainties. Virgo is dominated by quiescent galaxies through most of the stellar mass range, and particularly at low masses. At high masses, there are few SF galaxies (the SF fraction peaks around $10^{9.5}~\text{M}_{\odot}$), but larger abundances of low-SFR galaxies. This is also evident on Fig.~\ref{fig:SFMS}, where many of the most massive galaxies have some \ha{} emission but are well below the SFMS.

\subsection{The stellar mass function across the Virgo environment} \label{sec:env}

Because Virgo has a complicated structure, it is interesting to measure and compare the SMFs in different regions of the cluster with each other and with the field. First, we are interested in pure samples of galaxies associated with Cluster A. Using the Cluster A sample as defined in Section~\ref{sec:sub_scale}, we construct three bins based on distance to M87: $0~\text{Mpc} \leq R < 0.5~\text{Mpc}$, $0.5~\text{Mpc} \leq R < 1.0~\text{Mpc}$ and $1.0~\text{Mpc} \leq R < 1.5~\text{Mpc}$ (Core, Inner Ring, and Outer Ring, respectively). Adopting $R_{200}\sim 1~\text{Mpc}$ \citep{simionescu2017} as the virial radius of Virgo, we can consider galaxies in the Outer Ring to be in the infall region of the main cluster.

\begin{figure*}[ht!]
\centering
\includegraphics[width=2\columnwidth]{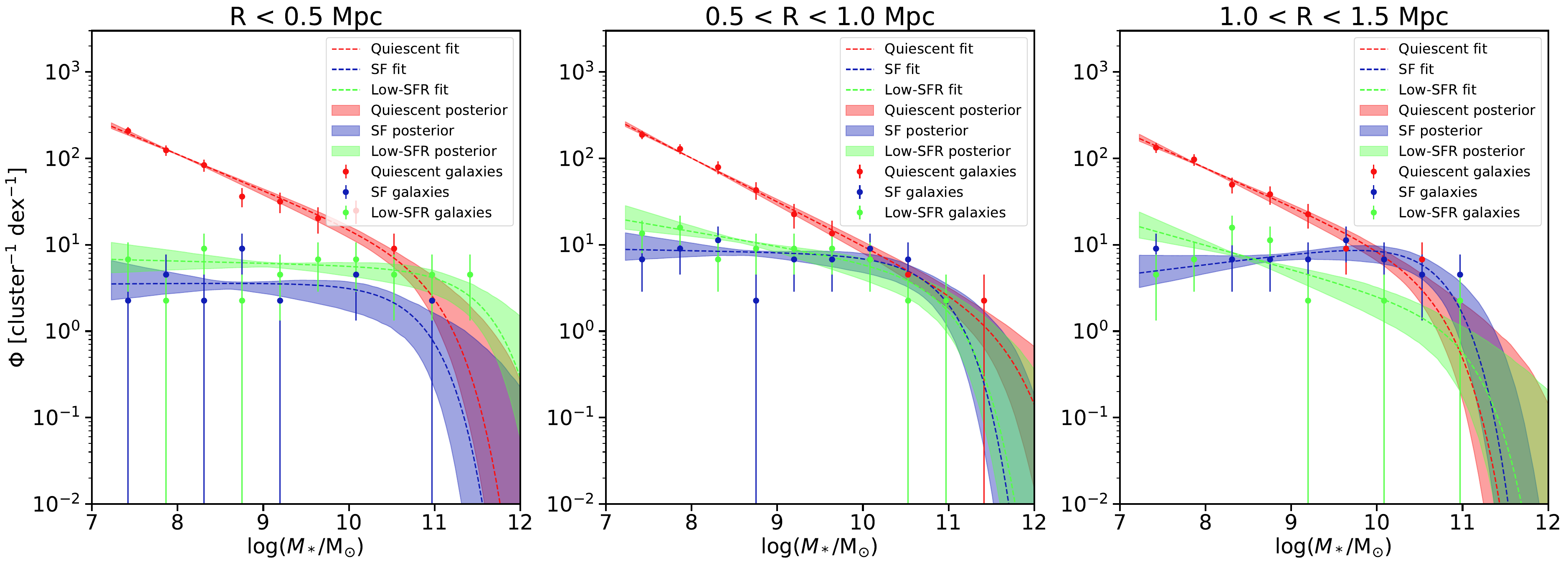}
\caption{SMFs for the quiescent, low-SFR and SF populations in three radial bins centred on M87. Methodology and colours are the same as Figure \ref{fig:smf_sfms1}.}
\label{fig:smf_env}
\end{figure*}

\begin{figure*}[ht!]
\centering
\includegraphics[width=2\columnwidth]{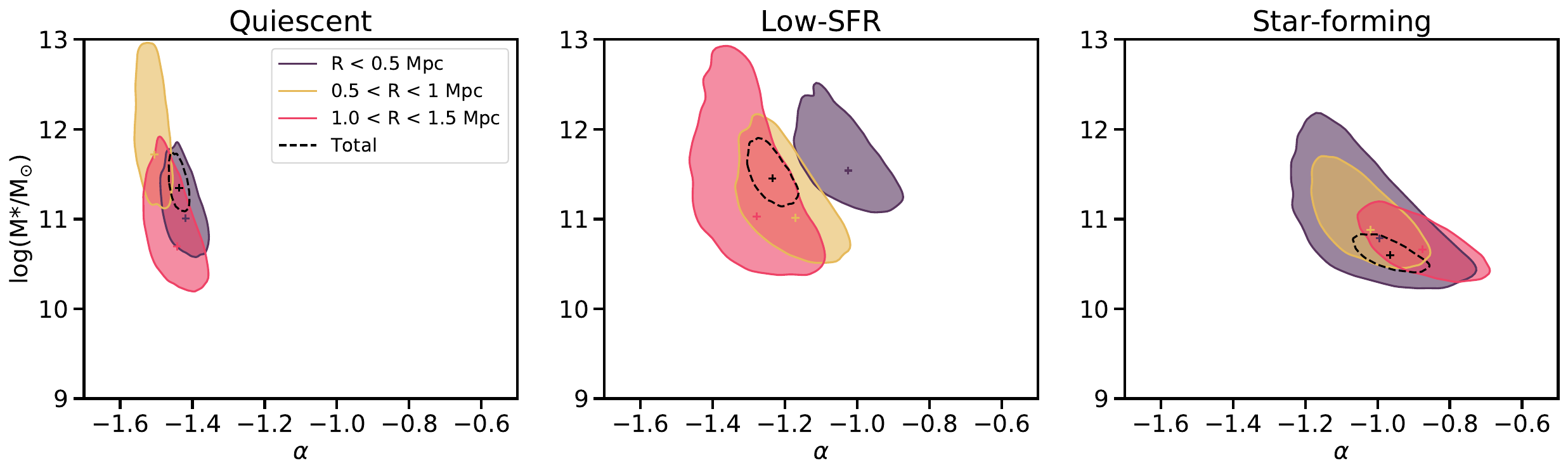}
\caption{Best fit and $1\sigma$ contours for the fit values of $\alpha$ and $\mathcal{M}^*=\log{\left(M^*/\text{M}_{\odot}   \right)}$ for quiescent, low-SFR and SF SMFs, shown for the three radial bins of Cluster A and the total Virgo population. The black data point and contour represent the best fit and $1\sigma$ for the total SMF of all Virgo galaxies. Errorbars on the fractions display binomial confidence intervals.}
\label{fig:params_env}
\end{figure*}

We show in Figure \ref{fig:smf_env} the SMF fits for quiescent, low-SFR and SF populations in the three Cluster A radial bins. Since we have constructed a pure sample of Cluster A galaxies located at 16.5~Mpc, we can safely fit these SMFs down to $\mathcal{M}_{lim} > 7.2$.  In Figure \ref{fig:params_env} we show the best fit and $1\sigma$ contours for the two free Schechter parameters ($\alpha$ and $\mathcal{M}^*=\log{(M^*/\text{M}_{\odot})}$) for the three populations in each of these bins as well as the total Virgo population. The shapes of the mass functions differ only slightly and insignificantly across the three radial bins. In each case, the population is dominated by a steep power-law quiescent mass function and comparatively flat SF mass functions. The low-SFR SMF has a slope that is intermediate between the SF and quiescent populations, indicating that these galaxies are not simply a subset of one of the other populations. There are no significant environmental variations in the characteristic masses of each population.

\begin{figure*}[ht!]
\centering
\includegraphics[width=2\columnwidth]{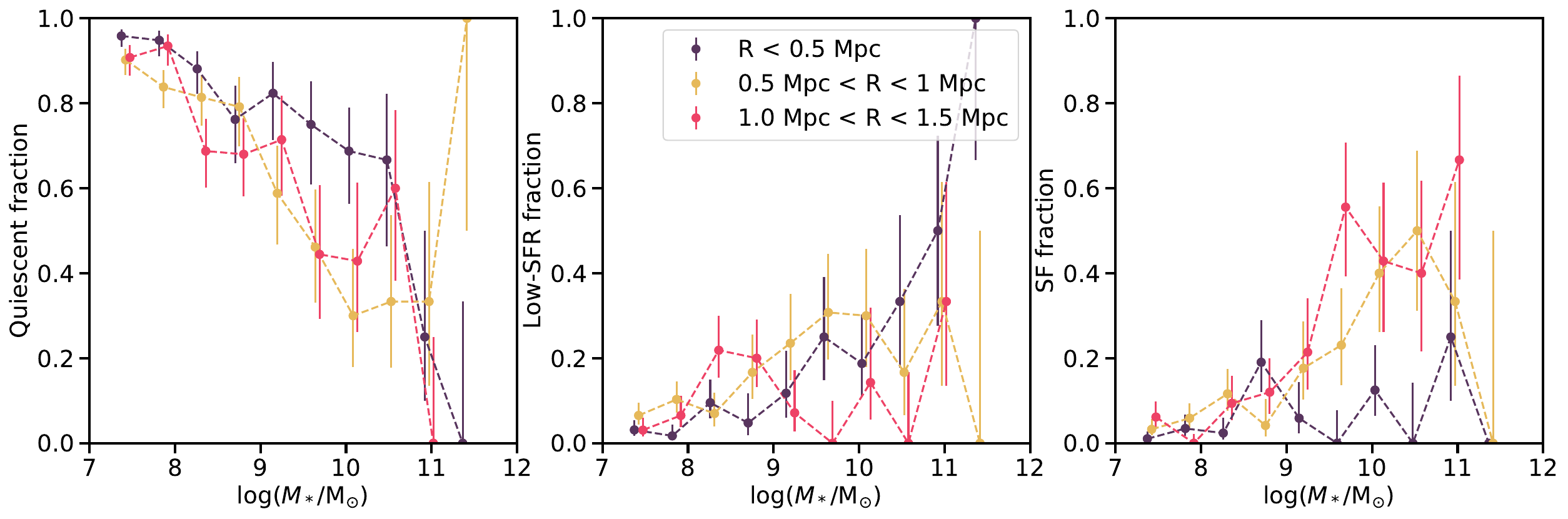}
\caption{Quiescent, low-SFR and SF fractions as a function of stellar mass. Fractions are shown for the three radial bins of Cluster A based on the binned stellar masses shown in Figures \ref{fig:smf_env} and \ref{fig:smf_env2}.}
\label{fig:frac_env}
\end{figure*}

\clearpage

In Figure \ref{fig:frac_env} we show the fractions of quiescent, low-SFR and SF galaxies as a function of stellar mass in the three Cluster A radial bins. Across the Cluster A environment, the low-mass end is dominated by quiescent galaxies. As one moves toward the high-mass end, the quiescent fractions decrease, and the low-SFR and SF fractions increase. Quiescent and low-SFR galaxies tend to dominate the high-mass regime again beyond $\mathcal{M} > 11$, though low numbers of galaxies in this regime increase the uncertainties on these fractions. The SF fraction peaks in each environment at $\mathcal{M} 
\sim 10.5$. While the shapes of the SMFs do not change significantly across the environment, the relative contributions of the quiescent, SF and low-SFR populations do change, with slightly lower quiescent fractions (below $\mathcal{M} < 10.5$) in the Inner Ring and Outer Ring than the Core. The \ha{}-emitting population in the Outer Ring is dominated by SF galaxies, whereas in the Inner Ring, low-SFR galaxies dominate. With that said, the quiescent populations dominate across all environments in this mass range. We will further discuss the impact of the Virgo environment on the relative fraction of quiescent and low-SFR galaxies and compare to values seen in the field in Sect. \ref{sec:discussion}.

\section{Discussion} \label{sec:discussion}

\subsection{Cluster B and the Clouds}

\begin{figure*}[ht!]
\centering
\includegraphics[width=2\columnwidth]{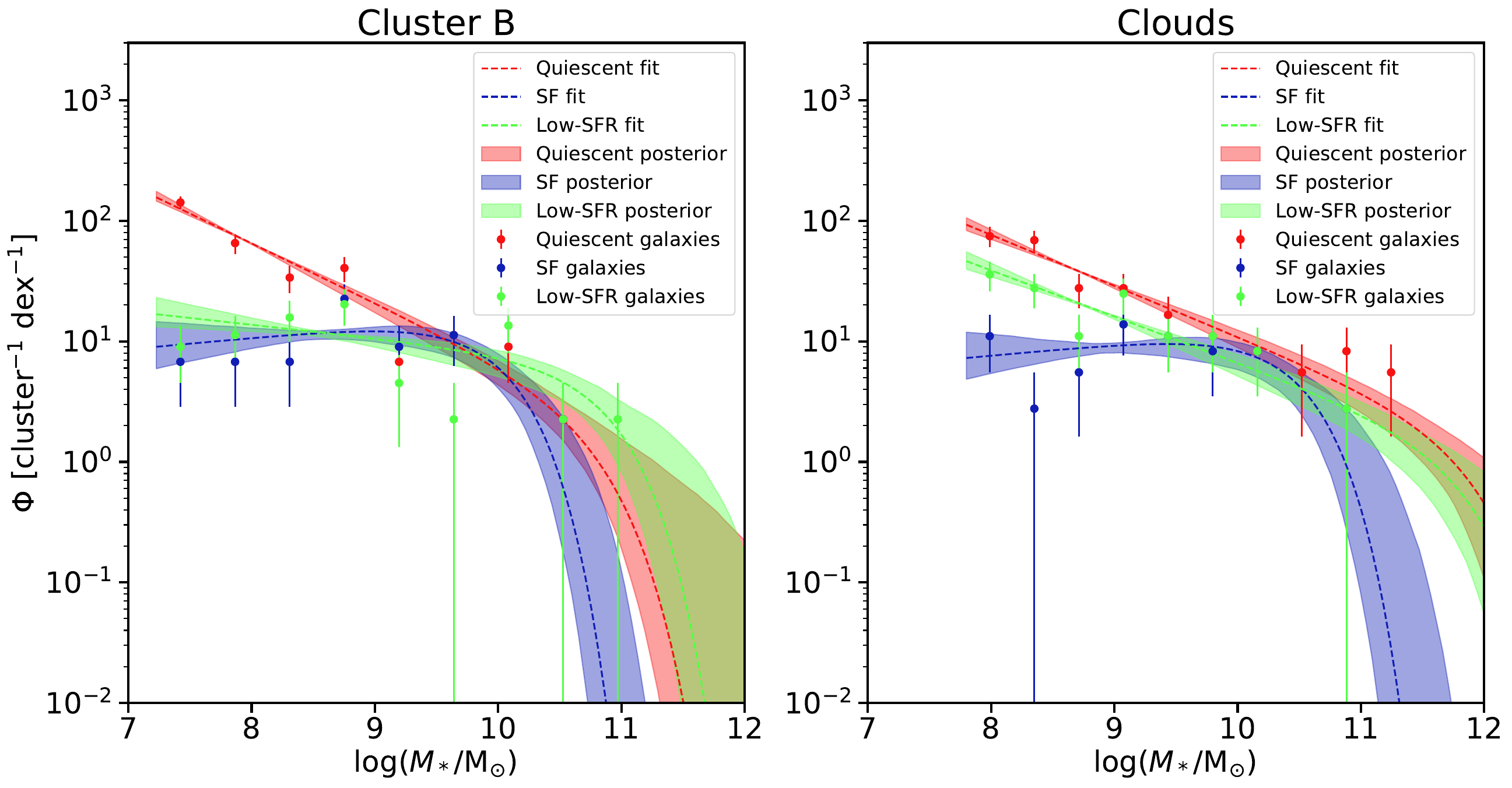}
\caption{SMFs for the quiescent, SF and low-SFR populations in Cluster B and the Clouds. Methodology and colours are the same as Figures \ref{fig:smf_sfms1} and \ref{fig:smf_env}.}
\label{fig:smf_env2}
\end{figure*}

\begin{figure*}[ht!]
\centering
\includegraphics[width=2\columnwidth]{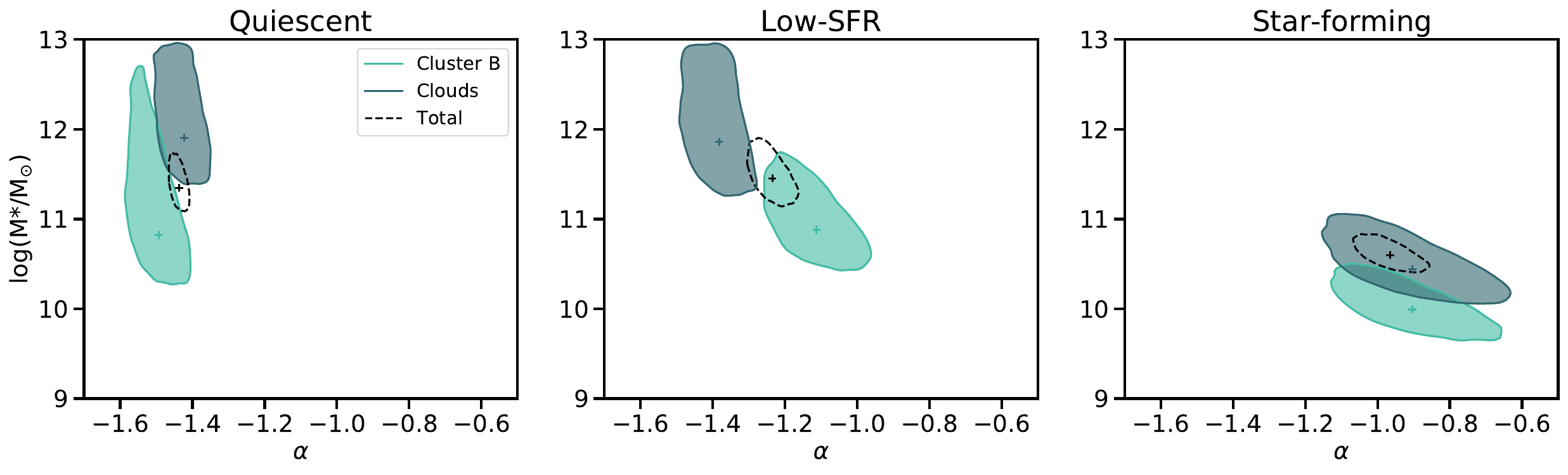}
\caption{Same as Figure~\ref{fig:params_env} but for the Cluster B and Clouds populations.}
\label{fig:params_env2}
\end{figure*}

\begin{figure*}[ht!]
\centering
\includegraphics[width=2\columnwidth]{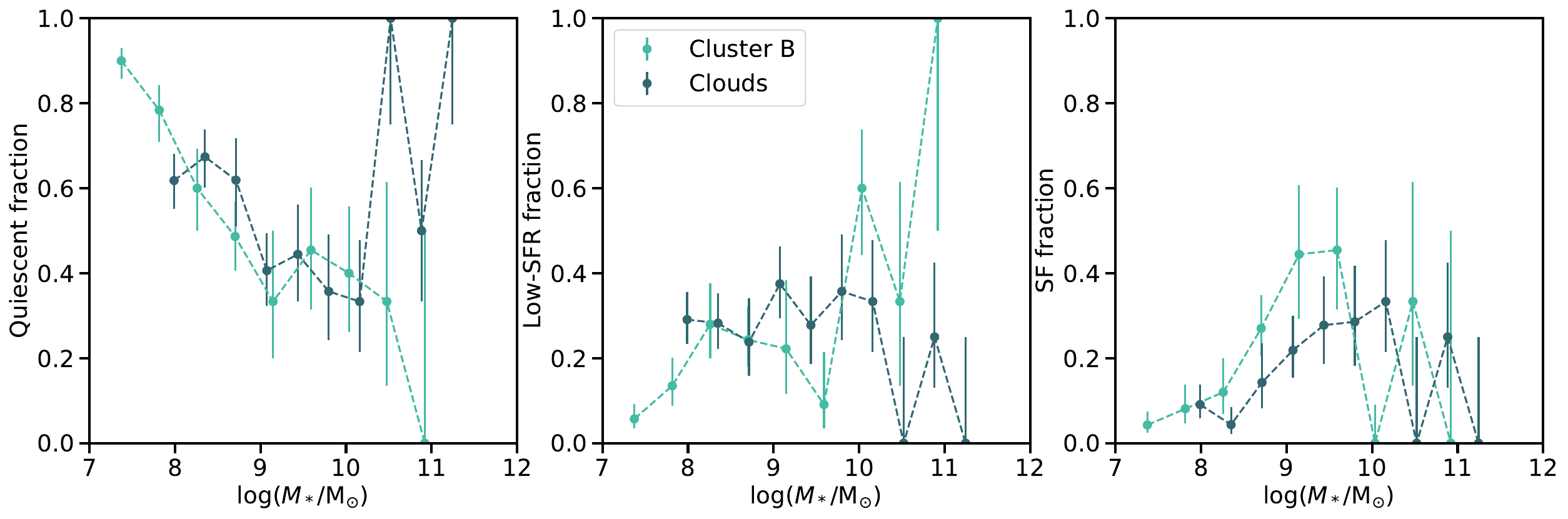}
\caption{Same as Figure~\ref{fig:frac_env} but for the Cluster B and Clouds populations.}
\label{fig:frac_env2}
\end{figure*}

When separating galaxies into different environmental bins, we assigned galaxies to Cluster B if they fell within the Cluster B footprint, and $>1~\text{Mpc}$ from M87. Galaxies that overlap in the virialized region of Cluster A and Cluster B are quite likely to have been affected by the Cluster A environment. In addition, there is no significant overdensity of galaxies in this region compared to the Outer Ring to assume they belong to Cluster B. We show the quiescent, low-SFR and SF SMF fits to the Cluster B and Clouds regions in Figure~\ref{fig:smf_env2}, with the fit parameters in Figure~\ref{fig:params_env2}. Since Cluster B is located at 16.5~Mpc, we fit the SMFs down to $\mathcal{M}_{lim} > 7.2$ in this region, while the Clouds SMFs can only be fit down to $\mathcal{M'}_{lim} > 7.78$ because of their greater distance. The SMFs of these two regions are again consistent with the total SMFs of the Virgo cluster within uncertainties. We note that the characteristic mass of the quiescent and low-SFR SMFs of the Clouds are higher than other regions. We do not expect the Clouds region to be pure - some galaxies may actually be located at 16.5~Mpc and be part of Cluster A, B or the LV Cloud. If some galaxies have incorrectly had their masses scaled, this may be pushing the SMF to higher characteristic masses. We show the relative fractions of these regions in Figure~\ref{fig:frac_env2}. While we see a more moderate quiescent fraction and higher SF fraction in the Clouds, the quiescent fraction is still significant. However, it is difficult to determine whether this is due to contamination from Cluster A and B galaxies, or due to environmental effects in the Clouds themselves. 

In Cluster B, our sample should be quite pure, with some possible contamination from Outer Ring galaxies from Cluster A. Nonetheless, this population is made up of galaxies beyond the virial radius of Cluster A. While we do see a lower quiescent fraction and higher SF fraction at intermediate masses in Cluster B, the low mass end is still dominated by quiescent galaxies and the high mass end by low-SFR galaxies. We will discuss these results further in Section~\ref{sec:lit} when comparing to other studies of field and cluster quenched fractions.

\clearpage

\subsection{The need for pre-processing} \label{sec:preprocess}

\begin{figure}[ht!]
\centering
\includegraphics[width=\columnwidth]{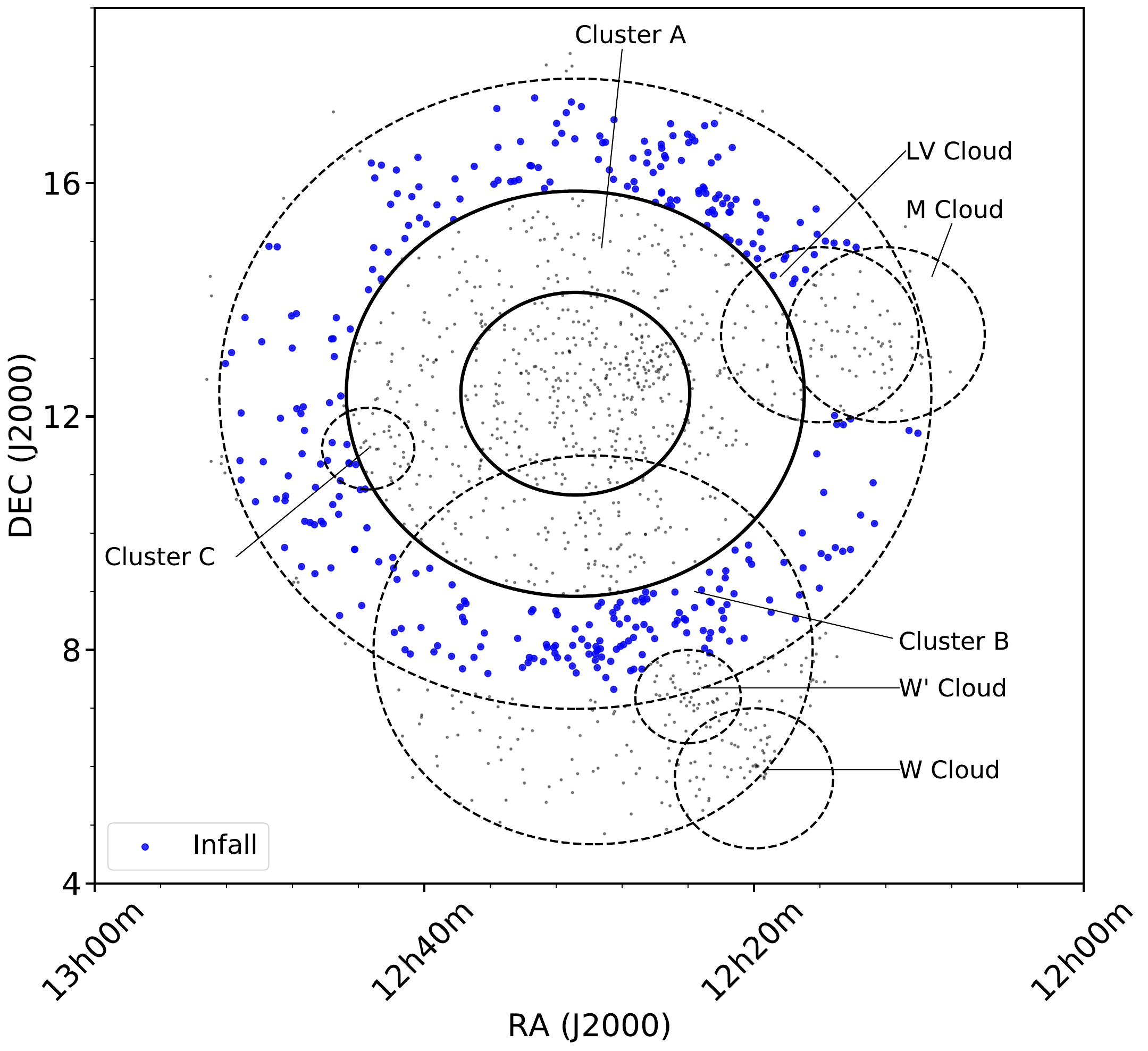}
\caption{Same as Fig.~\ref{fig:footprint} but highlighting just the Infall population.}
\label{fig:footprint_infall}
\end{figure}

In an attempt to explore the infall region of the cluster, we define a new Infall region that contains the Outer Ring as defined in the previous section, but also any Cluster~B galaxies that fall from $1.0~\text{Mpc} \leq R < 1.5~\text{Mpc}$ (we continue to exclude galaxies assigned to the W, W$^\prime$ and M clouds since they are located far behind the cluster). The choice to include Cluster~B galaxies stems from the fact that at $z=0$, $\sim~40\%$ of galaxies in clusters were satellites of group halos prior to cluster infall \cite[e.g.][]{mcgee2009}. Thus, a true infall population should include galaxies infalling as part of smaller halos. We show the galaxies being analyzed in Figure~\ref{fig:footprint_infall}.

Fig.~\ref{fig:smf_full} shows that the SMF of quiescent Virgo galaxies is of a completely different shape from that of the field.  This is true even of the Infall and Outer Ring populations, for which the SMF is statistically indistinguishable from the Core (Fig.~\ref{fig:params_env})\footnote{We note that we do not show the SMF for the quiescent Infall population as it is essentially identical to that of the Outer Ring population shown in Fig.~\ref{fig:smf_env}. For completeness, we run the following tests using both the Infall and Outer Ring populations and find no difference in the results.}.  Therefore, environmental effects must dominate the galaxy population, even in the Infall regions of Virgo. One possibility is that this region is dominated by backsplash galaxies. The similarity between the Core and Infall region SMFs means that the contribution from field quiescent galaxies must be small and backsplash galaxies would have to account for a large majority of the infall population, which is not supported by simulations \citep[e.g.][]{haggar2020}. The other possibility is that galaxies are being environmentally pre-processed in smaller halos before their infall into Virgo, or are being processed by Virgo well beyond $R_{200}$. By quantifying the maximum contribution from field quiescent galaxies allowed given the statistical uncertainties on the data, we can attempt to rule out the backsplash-only scenario in favour of a scenario involving some degree of pre-processing.

We consider a simple model where the quiescent SMF in the Infall bin is comprised of a linear combination of the field quiescent mass function and the Core quiescent mass function. This assumes that the environmentally processed quiescent galaxy population has a universal SMF shape, which is the simplest assumption given the lack of environmental dependence observed in the sample. It follows that

\begin{equation}
    \phi_{\rm q, infall} = A\times\phi_{\rm q, core} + B\times\phi_{\rm q, field},
\end{equation}

\noindent where $A$ and $B$ are coefficients that determine the relative contributions of the core and field populations. $\phi_{\rm q, core}$ is the quiescent population in the core of Virgo shown in Fig.~\ref{fig:smf_env}, while $\phi_{\rm q, field}$ represents the red population\footnote{While the definition of red galaxies from GAMA is based on colour, we noted previously that the work of \citealt{gunawardhana2015} found nearly a nearly identical SMF for \ha{}-selected GAMA galaxies as the blue population in \citealt{baldry2012}. This implies that a quiescent population in GAMA selected based on \ha{} non-detections would have a similar SMF to the red population, justifying the use of the GAMA red SMF here.} shown in Fig.~\ref{fig:smf_full}, taken from the GAMA survey \citep{baldry2012}. The fraction of galaxies coming from the core and field populations respectively can be determined by 

\begin{equation}
    f_{\rm q, core} = \frac{A\int_{M_{\rm lo}}\phi_{\rm q, core}}{\int_{M_{\rm lo}}\phi_{\rm q, outer}}    
\end{equation}    

\noindent and

\begin{equation}
    f_{\rm q, field} = \frac{B\int_{M_{\rm lo}}\phi_{\rm q, field}}{\int_{M_{\rm lo}}\phi_{\rm q, outer}},
\end{equation}  

\noindent where $M_{\rm{lo}}=10^{7.2}~\text{M}_{\odot}$ is the lower completeness limit for galaxies in the NGVS sample as outlined in Sect.~\ref{sec:NGVS}.

\begin{figure}[ht!]
\centering
\includegraphics[width=\columnwidth]{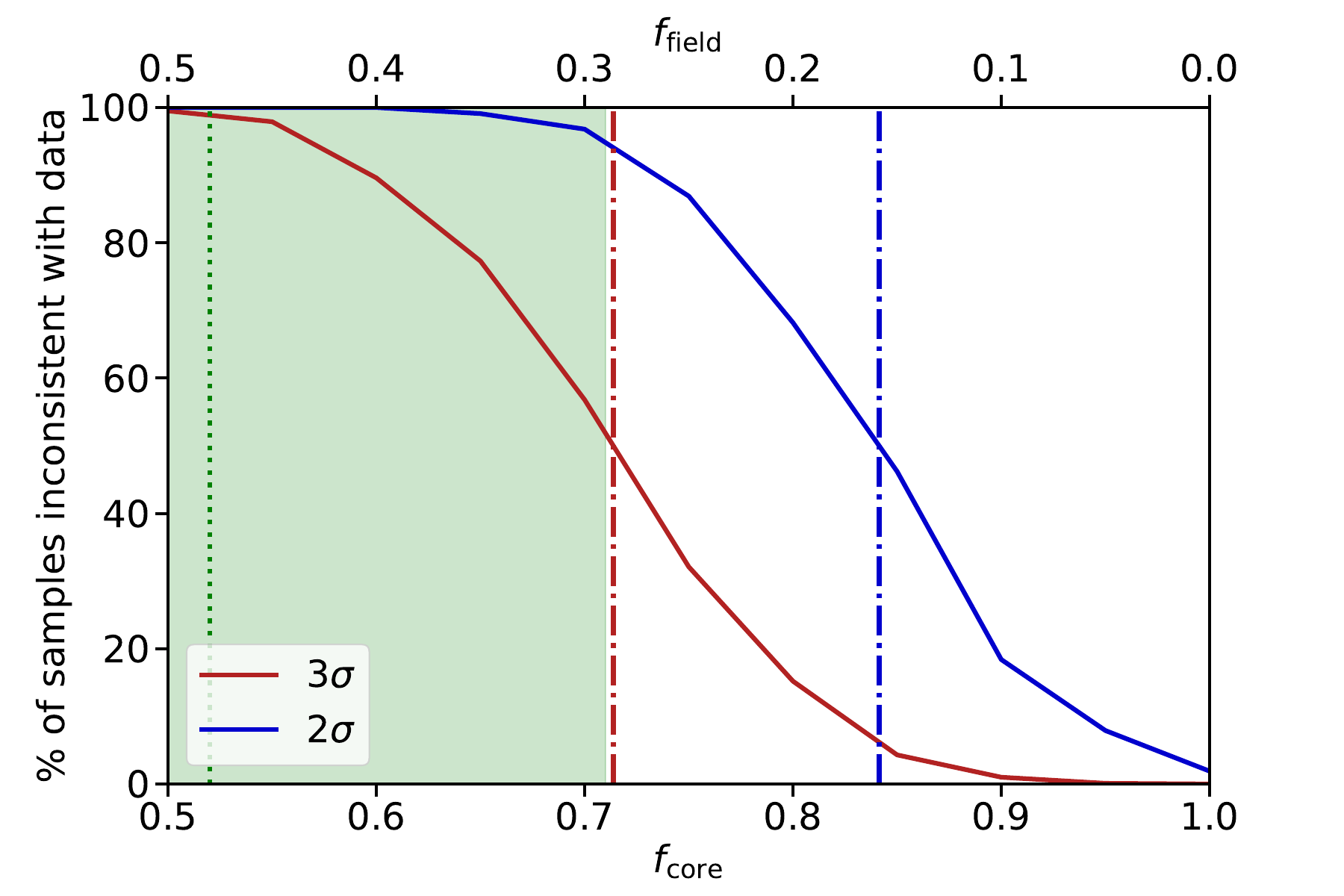}
\caption{Percentage of field+core quiescent samples that are inconsistent with the observed quiescent population in the Infall region of Virgo, as a function of core fraction. Samples are determined to be inconsistent at a $2\sigma$ (blue curve) or $3\sigma$ (red curve) level based on an Anderson-Darling test comparing the samples with the observed Infall population. The dashed-dot lines show where $<50\%$ of the samples are determined to be inconsistent with the data, meaning that the majority of samples drawn cannot be ruled out as being drawn from the same distribution as the data. The green dotted line shows the median backsplash fraction between $1$ and $1.5R_{200}$ from \citet{haggar2020} for unrelaxed clusters, while the shaded green area shows the 68\% confidence interval . This test shows that the maximum field galaxy contribution to the Infall quiescent population is not enough to remedy the high backsplash fractions, implying the need for pre-processing beyond $R_{200}$.}
\label{fig:AD}
\end{figure}

Since the quiescent SMFs in the Infall region and Core look similar, we expect a high fraction of Core galaxies. As such, we vary the Core fraction $f_{\rm q, core}$ between 0.5 and 1, with $f_{\rm q, field}=1-f_{\rm q, core}$. We then draw a sample of 247 galaxies (the number of observed quiescent galaxies in the Infall region), split between the Core and field populations based on $f_{\rm q, core}$ and $f_{\rm q, field}$, and run an Anderson-Darling (AD) test between this population and the real population of galaxies in the Infall region. For each value of $f_{\rm core}$, we repeat this test 1000 times to visualize the distribution of p-values from the AD tests. From the 1000 runs, we calculate the percentage of samples that can be ruled out as inconsistent with the observed population at the $2\sigma$ and $3\sigma$ levels (p-values of 0.05 and 0.003, respectively), as shown in Figure~\ref{fig:AD}. The regions where the majority of our samples are consistent with the observed population are indicated as anything to the right of the blue and red dashed vertical lines. When using a $2\sigma$ AD confidence, the majority of drawn samples look like the observed population for $f_{\rm core} > 0.84$. This drops to $f_{\rm core} > 0.71$ if we use a $3\sigma$ confidence level for the AD tests. In both cases, the Core fraction limits show that the fraction of environmentally-impacted galaxies in the Infall region is much higher than the fraction of backsplash galaxies at this radius found in simulations \citep{haggar2020}, $52\pm19\%$, shown as the green shaded region in Figure~\ref{fig:AD}.

Assuming that Virgo has a typical fraction of backsplash galaxies ($\sim52\%$), this rules out the hypothesis that all environmentally-quiescent galaxies in the Infall region are backsplash. Thus, there must be some amount of pre-processing in at least $\sim 20$ per cent of the population, prior to the first crossing of $R_{200}$ into the main Virgo cluster. There is significant cluster-to-cluster variation in the backsplash fractions determined from simulations. If Virgo has a backsplash fraction of $>1\sigma$ above the median, the conclusions become less significant, as a majority of drawn samples cannot be ruled out as different from the Infall population at the $3\sigma$ level.

Thus, pre-processing likely plays an important role in shaping the quiescent mass functions in the Virgo cluster. This inference is bolstered by the observed quiescent SMF in Cluster~B and the Clouds. The Clouds are located behind the main cluster, meaning that if galaxies in this region are quiescent, they have either quenched secularly in the field, or quenched via environmental mechanisms in the much smaller group environments. Processes such as RPS are likely much less common in smaller groups, since small groups lack the intracluster medium density of massive clusters. However, some cases of RPS in group-sized halos have been identified \citep[e.g.][]{catinella2013, roberts2021, kolcu2022}. Additionally, while direct stripping of cold gas is difficult, the loosely bound circumgalactic medium (CGM) could be stripped, driving starvation if the galaxy cannot accrete fresh gas. Given that dwarf galaxies will have smaller restoring forces with which to hold onto their gas, stripping of dwarf galaxies in group structures in more likely than stripping of massive spirals. Since most galaxies in the local universe are found in groups \citep[e.g.][]{geller1983, robotham2011}, pre-processing in these structures likely plays an important role in building up the quiescent population in Virgo. While this holds true for the Virgo cluster, results may be different for other local clusters, especially those are are dynamically inactive.

In addition, one cannot rule out the effects of the main cluster environment beyond the virial radius. The intracluster medium may extend far enough to strip gas from dwarf galaxies \citep[e.g.][]{boselli2022}, and starvation may begin well before the first crossing of $R_{200}$ \citep[e.g.][]{bahe2013,morgan2024}. It is also worth noting that various measurements of $R_{200}$ exist for Virgo, with values as high as 1.7~Mpc \citep{kashibadze2020}. The variation in $R_{200}$ is due in part to the asymmetric structure of the cluster, where different studies have made different choices about how to treat the southern extension of Virgo (i.e. Cluster B). If the virial radius is in fact much larger than the value we have adopted here, the Infall region as described would actually be within the virial radius.

\newpage

\subsection{Literature comparison} \label{sec:lit}

\subsubsection{Cluster and field Schechter fits}

\begin{figure}[ht!]
\centering
\includegraphics[width=\columnwidth]{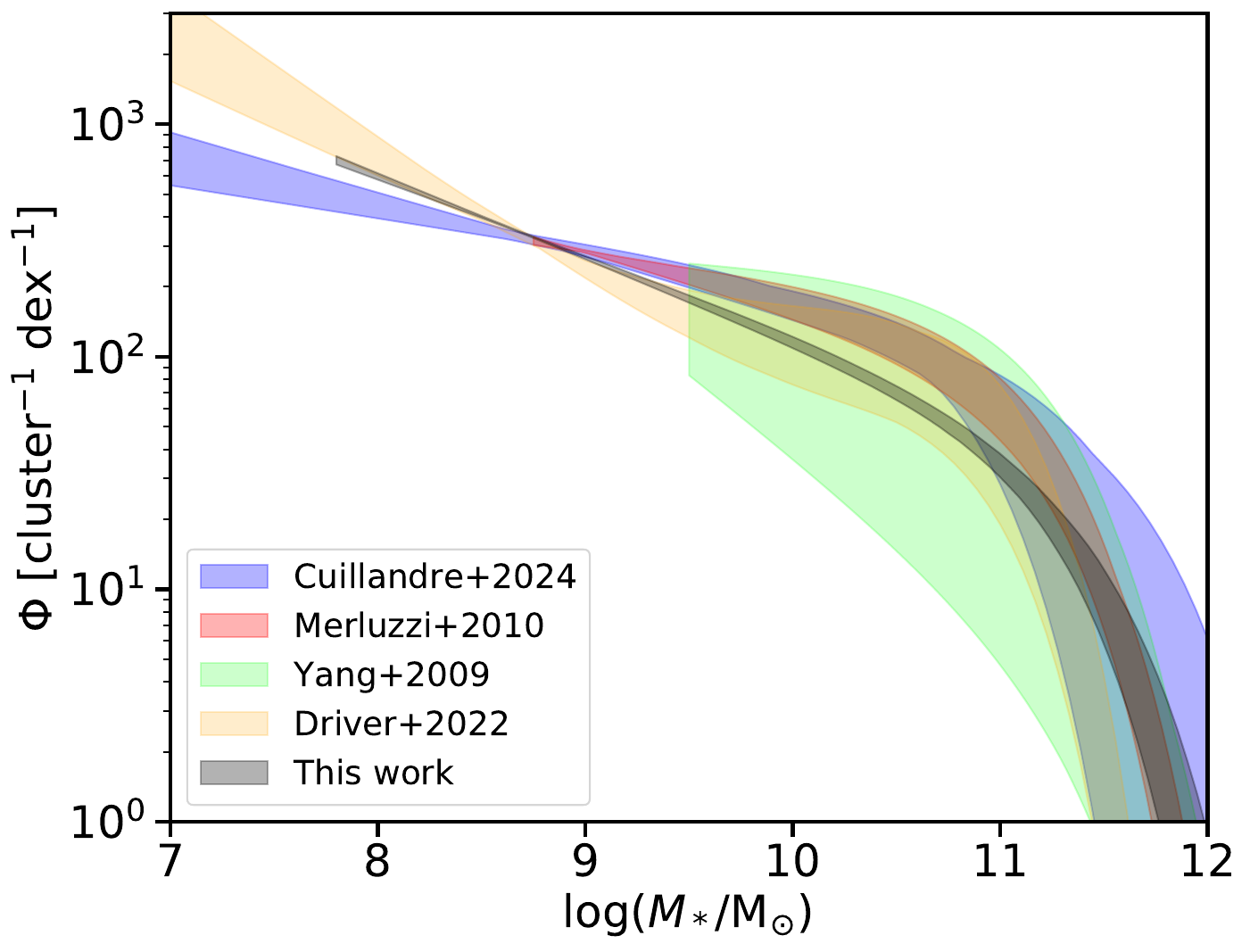}
\caption{Schechter fits for our total Virgo SMF compared to three other cluster samples, as well as the GAMA field fit from \citet{driver2022}.}
\label{fig:params_lim}
\end{figure}

\citet{ferrarese2016} determined the total SMF in the very core of Virgo (central $0.3~\text{Mpc}^2$) to have a slope of $\alpha\sim-1.35$, though their fit did not effectively constrain the characteristic mass. Combining our quiescent, SF and low-SFR populations into a total Core population yields a SMF with $\alpha=-1.36$, consistent with the results of \citet{ferrarese2016}. Our Schechter fit for the total SMF of Virgo yields $\alpha=-1.35$ and $\mathcal{M}^*=11.33$, which we compare with other cluster Schechter fits in Figure~\ref{fig:params_lim}. The comparison studies include the Perseus cluster \citep{cuillandre2024}, the Shapley Supercluster \citep{merluzzi2010} and SDSS halos with similar mass to Virgo ($10^{13.8} < M_h/\text{M}_{\odot} < 10^{14.4}$; \citealt{yang2009}). Our value for the low-mass slope $\alpha$ is consistent with the SDSS value and steeper than the Perseus and Shapley studies. \citet{cuillandre2024} note that a fit to the low-mass end of their Perseus sample gives a slope of $\alpha=-1.35$, consistent with the slope in our Virgo sample. Steeper low-mass slopes are consistent with deep field samples where a double-Schechter fit was used, such as in \citet{baldry2012}, where a low-mass slope of $\alpha=-1.47$ is found. 

While the slopes of the cluster and field samples are often consistent within their uncertainty, we find a steeper $\alpha$ than field values. In addition, the characteristic mass is always found to be higher in clusters than the field. While our characteristic mass is consistent with the cluster fits we compare to, typical field values are often smaller by 0.5-1~dex.

\subsubsection{Quiescent and low-SFR fractions}

While galaxies are often split into a bimodal population of star-forming and quiescent, either via a direct SFR tracer or by colour, we have isolated low-SFR galaxies from our sample to probe this transition population. We are able to isolate this sample by choosing galaxies with \ha{} emission from star formation that fall below the SFMS. As is seen in Fig.~\ref{fig:smf_sfms1}, the low-mass end of our sample is dominated by quiescent galaxies showing no evidence for ongoing star formation. The relative fraction of SF galaxies peaks near the middle of our stellar mass range, and the low-SFR fraction increases with increasing stellar mass.

\begin{figure}[ht!]
\centering
\includegraphics[width=\columnwidth]{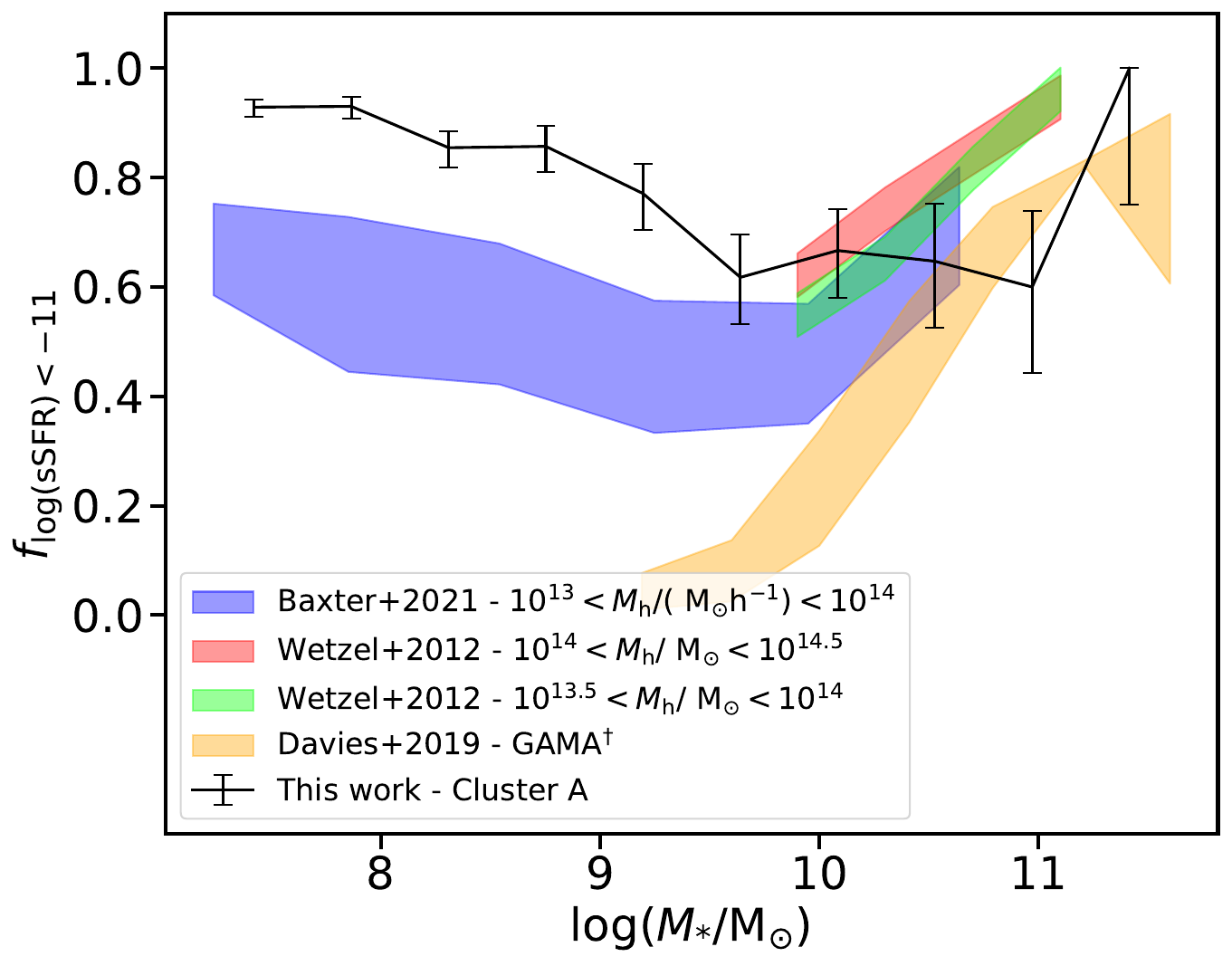}
\caption{Quiescent fraction as a function of stellar mass. The black data with error bars are Virgo galaxies from Cluster A, with the quiescent sample chosen as galaxies with $\log{(\rm{sSFR})}<-11$. The red and green shaded regions show $1\sigma$ uncertainty ranges for the quiescent fractions from \citet{wetzel2012} for two halo mass bins: $10^{14} < M_h/\text{M}_{\odot} < 10^{14.5}$ and $10^{13.5} < M_h/\text{M}_{\odot} < 10^{14}$. The blue shaded region shows the satellite quiescent fractions for the halo mass range $10^{13} < M_h/(\text{M}_{\odot}h^{-1}) < 10^{14}$ from \citet{baxter2021}. The purple region shows the range of quiescent ($\log{(\rm{sSFR})}<-10.5$) fractions for the GAMA sample from \citet{davies2019}, where the width of the GAMA shaded region is shown to encompass both the satellite and central galaxy quiescent fractions. \citet{wetzel2012} and \citet{baxter2021} use $\log{(\rm{sSFR})}<-11$ as their quiescent cut, while \citet{davies2019} uses $\log{(\rm{sSFR})}<-10.5$.}
\label{fig:qf_mstel}
\end{figure}

\begin{figure*}[ht!]
\centering
\includegraphics[width=2\columnwidth]{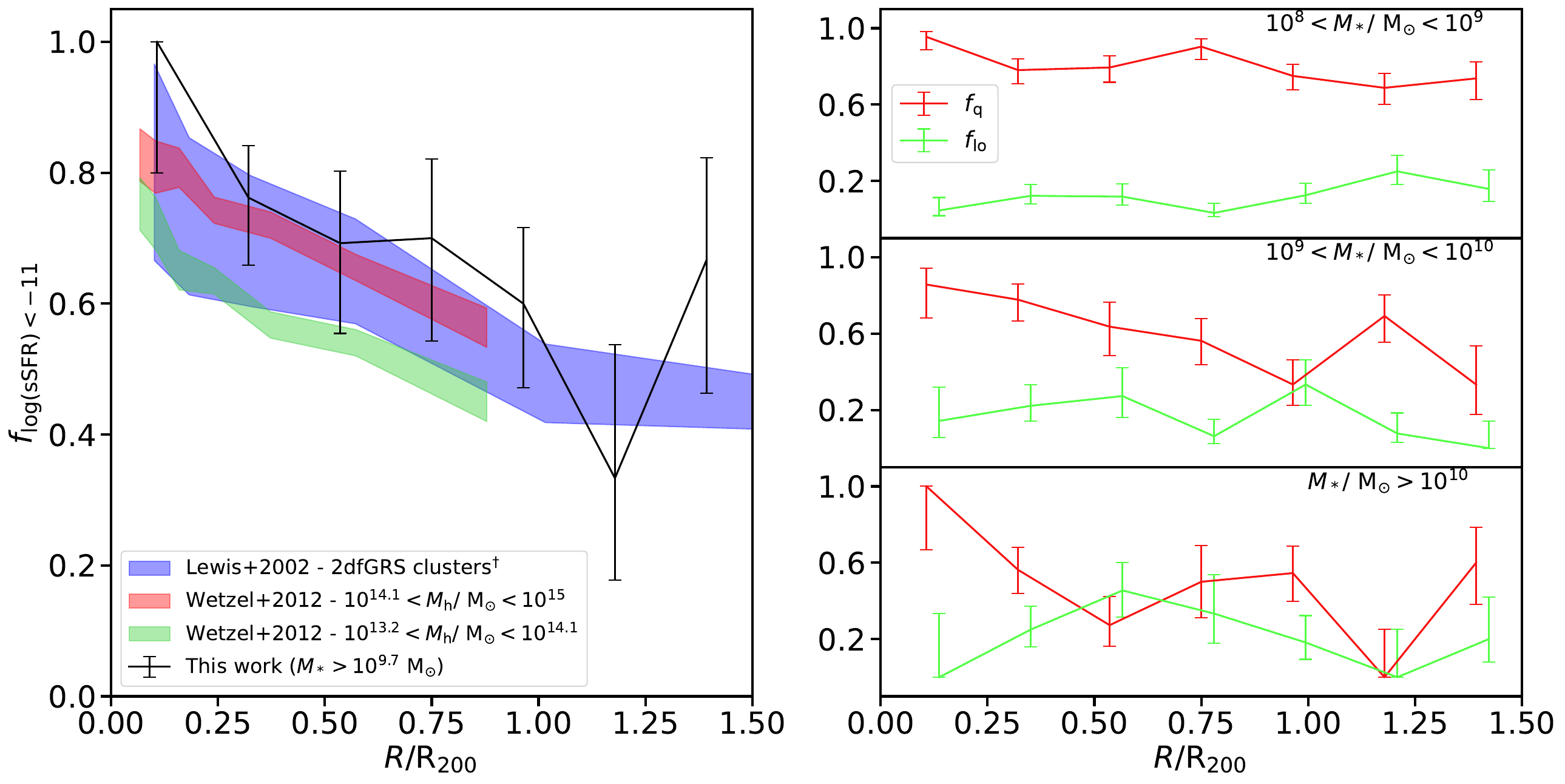}
\caption{Quiescent and low-SFR fractions as a function of clustercentric distance. \textit{Left:} The black data with error bars are for Virgo galaxies from Cluster A, with the quiescent sample chosen as galaxies with $\log{(\rm{sSFR})}<-11$, and have limited the stellar mass range to be $M_* > 10^{9.7}~\text{M}_{\odot}$ for consistency with comparison studies. The red and green shaded regions show $1\sigma$ uncertainty ranges for the quiescent fractions from \citet{wetzel2012} for two halo mass bins: $10^{14.1} < M_h/\text{M}_{\odot} < 10^{15}$ and $10^{13.2} < M_h/\text{M}_{\odot} < 10^{14.1}$. The blue shaded region shows the quiescent fractions from the cluster sample of the 2dFGRS \citep{lewis2002}. While the \citet{wetzel2012} study uses $\log{(\rm{sSFR})}<-11$ to isolate quiescent galaxies, the \citet{lewis2002} study uses a proxy for sSFR based on \ha{} equivalent width and continuum luminosity. \textit{Right:} Quiescent and low-SFR fractions based on our definitions from Section~\ref{sec:sf_q}, for galaxies of Cluster~A split into three different bins of stellar mass.}
\label{fig:qf_rad}
\end{figure*}

To compare our quiescent fraction trends with other works, we first produce a sample of all galaxies belonging to Cluster A, the main cluster of Virgo (Core, Inner Ring and Outer Ring). We then re-determine a quiescent sample by taking all galaxies with $\log{(\rm{sSFR})}<-11$ for proper comparison with other works that employ a sSFR cut (where $\text{sSFR}=\text{SFR}/\text{M}_*$). We note that, because of the depth of our \ha{} data, this new quiescent sample will contain low-SFR galaxies at all stellar masses. We show the quiescent fraction of this sample as a function of stellar mass in Figure~\ref{fig:qf_mstel}. For comparison, we show the quiescent fractions from \citet{wetzel2012} for two halo mass bins ($10^{13.5} < M_h/\text{M}_{\odot} < 10^{14}$ and $10^{14} < M_h/\text{M}_{\odot} < 10^{14.5}$; the virial mass of Cluster~A is $\sim10^{14}~\text{M}_{\odot}$ based on \citealt{simionescu2017}). In addition we show the satellite quiescent fractions from \citet{baxter2021} for local massive groups with halo mass $10^{13} < M_h/(\text{M}_{\odot}h^{-1}) < 10^{14}$. Finally, we show quiescent fractions from the GAMA survey \citep{davies2019}, encompassing both satellite and central quenched fractions. At low stellar masses, we find our quiescent fraction is much higher than that of the field, where the field has dropped to nearly zero at $\sim10^{9}~\text{M}_{\odot}$. At stellar masses less than this, our galaxies are $>80\%$ quiescent. Studies such as \citet{geha2012} show that from $\sim10^7~\text{M}_{\odot}$ to $\sim10^9~\text{M}_{\odot}$, fully quiescent galaxies are essentially non-existent in isolated field samples. However, low mass galaxies in large, low-redshift groups display high quenched fractions as shown by the values from \citep{baxter2021}. This again may speak to the impact of pre-processing of low-mass galaxies prior to infall into massive clusters, as discussed in Section~\ref{sec:preprocess}.

The interpretation made in Section~\ref{sec:preprocess} were motivated by the high fraction of quiescent galaxies seen in the Infall region of Virgo. We show on the left side of Figure~\ref{fig:qf_rad} the quiescent fraction as a function of clustercentric radius compared to two relevant studies. The first is a study of clusters in the 2df Galaxy Redshift Survey \citep[2dFGRS,][]{colless2001,lewis2002}, where a proxy for sSFR was used to separate SF and quiescent galaxies. The second is again the SDSS study of \citet{wetzel2012}, where we choose the halo mass bins $10^{13.2} < M_h/\text{M}_{\odot} < 10^{14.1}$ and $10^{14.1} < M_h/\text{M}_{\odot} < 10^{15}$. To make our stellar mass range comparable to these two studies, we restrict our sample only to galaxies with $M_*>10^{9.7}~\text{M}_{\odot}$ and choose only galaxies with $\log{(\rm{sSFR})}<-11$. We find that for more massive galaxies, our quiescent fractions are consistent with these previous studies even beyond $R_{\rm 200}$. It is low-mass galaxies, then, that are driving the high quiescent fractions in the outer regions of the Virgo cluster. 

On the right side of Fig.~\ref{fig:qf_rad}, we return to our original definition of quiescent and low-SFR galaxies and show the relative fractions of each in three bins of stellar mass as a function of clustercentric radius. For the two smaller stellar mass bins, we see less of a radial decline in quiescent fraction, indicating again that galaxies with lower stellar masses are likely quenching before first infall. Throughout the cluster we observe a typical low-SFR fraction of $\sim 10-20\%$, with values in certain radial bins as low as $\sim 0\%$ and as high as $\sim 50\%$. If this population is dominated by transition galaxies, then galaxies undergoing quenching can be found across the Virgo environment.

\newpage 
\section{Summary and conclusions} \label{sec:conclusions}

Using a deep stellar mass catalogue of the Virgo cluster environment constructed with NGVS optical data, we 
have produced and analyzed SMFs across the Virgo environment. With deep \ha{} data from VESTIGE, we split our galaxies into quiescent, low-SFR and SF samples, and compared the shapes of the SMFs across the Virgo environment. In general, the Virgo cluster is dominated by a quiescent population with steep low-mass slope while the SMFs of the low-SFR and SF populations are flatter. There are only marginal differences in the shapes of the SMFs across the Virgo environment. We summarize our key findings:

\begin{itemize}

    \item The shape of the total SMF of Virgo is similar to other cluster studies. The low-mass slope is similar to the field, with a larger characteristic mass (Fig.~\ref{fig:params_lim}). The shapes of the quiescent and SF SMFs in Virgo are significantly different from the field. The Virgo quiescent SMF has a nearly identical slope to the field SF SMF, consistent with the interpretation that the quiescent cluster population results from the quenching of SF field galaxies.
    
    \item The main body of Virgo, Cluster~A, is dominated by low-mass quiescent galaxies ($f_q > 0.7$ for galaxies with $M_*<10^9~\text{M}_{\odot}$), with a smaller number of more massive SF and low-SFR galaxies. This trend remains true in smaller substructures of Virgo such as the Clouds and Cluster~B, though the relative fraction of quiescent galaxies is as low as $\sim0.4$ in some mass bins (Figs.~\ref{fig:smf_env}-\ref{fig:frac_env2}).
    
    \item The Infall region of the Virgo cluster is largely dominated by quiescent galaxies. If one assumes that this population is made up of backsplash galaxies and field galaxies on their first infall, the backsplash fractions necessary to reproduce the observed population are $>70\%$, while simulation studies such as \citet{haggar2020} have shown that typical backsplash fractions in this region for dynamically active clusters are $\sim52\%$ (Fig.~\ref{fig:AD}). This result suggests galaxies may be pre-processed outside of the Virgo cluster virial radius, either by the cluster or by group structures prior to cluster infall.

    \item For $M_*<10^{9}~\text{M}_{\odot}$, the quiescent fractions ($\log{(\rm{sSFR})}<-11$) as a function of stellar mass are $>80\%$ in Cluster~A This is in stark contrast to isolated field galaxies ($0\%$), and is even higher than satellites of massive groups ($<75\%$; Fig.~\ref{fig:qf_mstel}). This again implies that low-mass galaxies are likely quenched partly or entirely before entering the cluster environment.

    \item For high-stellar mass galaxies, the quiescent fractions ($\log{(\rm{sSFR})}<-11$) in Cluster~A as a function of clustercentric distance are $\sim60\%$, consistent with studies of other massive systems at low redshift such as \citet{lewis2002} and \citet{wetzel2012}. At lower stellar masses, the quiescent fractions are $>70\%$, at all clustercentric distances. In addition, there is a population of low-SFR galaxies at all stellar masses and all clustercentric radii, implying that across the cluster, there are galaxies that may be undergoing quenching of their star formation.

\end{itemize}

Although Virgo is a dynamically young cluster with a mass of only $\sim10^{14}~\text{M}_{\odot}$, it is dominated by quiescent galaxies lacking star-formation, particularly at low stellar masses. This population is ubiquitous throughout the cluster and its various substructures, and is in stark contrast with the population of isolated, field galaxies. We infer that galaxy quenching is very efficient in dense environments at $z=0$, and likely begins long before galaxies reach the virialized body of the main cluster.

\section*{Acknowledgments}

We thank the referee for their comments which helped us to make improvements to this manuscript.\\

We are grateful to Roan Haggar, Guillaume Hewitt and Cam Lawlor-Forsyth for productive discussions regarding SMFs, quiescent fractions and backsplash galaxies that helped to tie this work together.\\

CRM acknowledges support from an Ontario Graduate Scholarship. MLB acknowledges support from an NSERC Discovery Grant. IDR acknowledges support from the Banting Fellowship program.
\\

The preparation of this manuscript and much of the research behind it took place at the University of Waterloo, on the traditional territory of the Neutral, Anishinaabeg and Haudenosaunee peoples. The campus is located on the Haldimand Tract, the land granted to the Six Nations that includes six miles on each side of the Grand River. The authors from the University of Waterloo community acknowledge the privilege to work on these traditional lands. We also acknowledge that the data for this research was collected at CFHT on the summit of Mauna Kea, a location with great cultural significance to the Native Hawaiian community.  \\

\indent \textit{Software:} \textsc{astropy}  \citep{astropy, astropy2, astropy3}, \textsc{emcee} \citep{emcee}, \textsc{ipython} \citep{ipython}, \textsc{matplotlib} \citep{matplotlib}, \textsc{numpy} \citep{numpy}, \textsc{pandas} \citep{pandas, pandasv1.0.1}, \textsc{scipy} \citep{scipy}, \textsc{topcat} \citep{topcat}.

\vspace{5mm}
\textit{Facilities:} Canada-France-Hawaii Telescope (CFHT)

\newpage

\bibliography{main}{}
\bibliographystyle{aasjournal}

\appendix

\section{Data table}
\label{appendix}
\FloatBarrier

\begin{deluxetable*}{ccccccc}
\caption{Best fit parameters and uncertainties for the fits to the quiescent, low-SFR and SF populations across the Virgo environment}
\tablehead{\colhead{Environment}&\colhead{Dist from M87}&\colhead{Sample}&\colhead{N}&\colhead{$\alpha$}&\colhead{$\mathcal{M*}$}&\colhead{$\phi*$}\\[-0.20cm]
\colhead{}&\colhead{(Mpc)}&\colhead{}&\colhead{}&\colhead{}&\colhead{$(\log{(M_*/\rm{M}_{\odot})})$}&\colhead{$(\rm{cluster}^{-1}\rm{dex}^{-1})$}}\label{tab:sfms_params_full}
\startdata
 &  & Quiescent & 239 & $-1.42^{+0.03}_{-0.05}$ & $11.01^{+0.78}_{-0.18}$ & $2.66^{+0.85}_{-1.70}$ \\
Cluster A - core & $R < 0.5$ & Low-SFR & 23 & $-1.03^{+0.08}_{-0.11}$ & $11.54^{+0.64}_{-0.24}$ & $2.29^{+0.98}_{-1.30}$ \\
 &  & Star-forming & 12 & $-0.99^{+0.11}_{-0.19}$ & $10.78^{+1.12}_{-0.25}$ & $1.61^{+0.93}_{-1.20}$ \\
\hline
 &  & Quiescent & 217 & $-1.51^{+0.03}_{-0.04}$ & $11.73^{+0.89}_{-0.32}$ & $0.57^{+0.37}_{-0.40}$ \\
Cluster A - inner ring & $0.5 < R < 1.0$ & Low-SFR & 33 & $-1.17^{+0.07}_{-0.12}$ & $11.02^{+0.94}_{-0.23}$ & $1.89^{+0.98}_{-1.34}$ \\
 &  & Star-forming & 27 & $-1.02^{+0.08}_{-0.13}$ & $10.89^{+0.62}_{-0.21}$ & $3.24^{+1.26}_{-1.94}$ \\
\hline
 &  & Quiescent & 161 & $-1.44^{+0.04}_{-0.07}$ & $10.70^{+1.07}_{-0.18}$ & $2.14^{+0.82}_{-1.62}$ \\
Cluster A - outer ring & $1.0 < R < 1.5$ & Low-SFR & 20 & $-1.28^{+0.08}_{-0.16}$ & $11.03^{+1.32}_{-0.30}$ & $0.62^{+0.46}_{-0.53}$ \\
 &  & Star-forming & 25 & $-0.88^{+0.10}_{-0.13}$ & $10.66^{+0.42}_{-0.19}$ & $5.48^{+2.17}_{-2.62}$ \\
\hline
 &  & Quiescent & 138 & $-1.49^{+0.04}_{-0.07}$ & $10.82^{+1.38}_{-0.20}$ & $1.16^{+0.49}_{-1.00}$ \\
Cluster B & $R > 1.0$ & Low-SFR & 36 & $-1.11^{+0.06}_{-0.12}$ & $10.88^{+0.71}_{-0.20}$ & $2.84^{+0.87}_{-1.82}$ \\
 &  & Star-forming & 29 & $-0.90^{+0.13}_{-0.18}$ & $9.99^{+0.40}_{-0.18}$ & $7.25^{+3.06}_{-3.52}$ \\
\hline
 &  & Quiescent & 92 & $-1.42^{+0.05}_{-0.06}$ & $11.91^{+0.73}_{-0.33}$ & $0.74^{+0.49}_{-0.48}$ \\
Clouds & ... & Low-SFR & 49 & $-1.38^{+0.07}_{-0.08}$ & $11.86^{+0.76}_{-0.39}$ & $0.56^{+0.45}_{-0.39}$ \\
 &  & Star-forming & 23 & $-0.90^{+0.13}_{-0.21}$ & $10.44^{+0.51}_{-0.19}$ & $5.73^{+2.59}_{-3.45}$ \\
\hline
 &  & Quiescent & 247 & $-1.42^{+0.04}_{-0.06}$ & $10.51^{+0.52}_{-0.17}$ & $4.53^{+1.74}_{-2.53}$ \\
Infall & $1.0 < R < 1.5$ & Low-SFR & 42 & $-1.20^{+0.06}_{-0.11}$ & $11.04^{+0.91}_{-0.22}$ & $1.97^{+0.99}_{-1.36}$ \\
 &  & Star-forming & 39 & $-0.88^{+0.09}_{-0.10}$ & $10.55^{+0.30}_{-0.16}$ & $8.75^{+2.96}_{-2.98}$
\enddata
\end{deluxetable*}

\end{document}